\documentclass[journal,comsoc]{IEEEtran}
\IEEEoverridecommandlockouts
\usepackage{epstopdf}
\usepackage{epsfig}
\usepackage{cite}
\usepackage{amsmath,amsthm,amssymb,amsfonts}
\usepackage{algpseudocode}
\usepackage{algorithm}
\usepackage{graphicx}
\usepackage[hyphens]{url}
\usepackage[T1]{fontenc}
\usepackage{ulem}
\usepackage[font=small,labelfont=bf]{caption}
\usepackage{hyperref}
\usepackage{booktabs}
\usepackage{textcomp}
\usepackage{color}
\usepackage{amsmath,bm}
\usepackage{makecell}
\usepackage{comment}
\usepackage{subfig}
\usepackage[switch]{lineno}
\allowdisplaybreaks[4]

\usepackage{textcomp}
\usepackage{xcolor}
\newtheorem{proposition}{Proposition}
\newtheorem{definition}{Definition}
\newtheorem{theorem}{Theorem}

\newtheorem{case}{Case}

\newcommand{\NAME}{\textsc{NetStorm}}
\newcommand{\new}{\textcolor{black}}

\begin{document}

\title{Accelerating Geo-distributed Machine Learning with Network-Aware Adaptive Tree and Auxiliary Route}

\author{
Zonghang Li, 
Wenjiao Feng, 
Weibo Cai, 
Hongfang Yu, 
Long Luo, 
Gang Sun,
\IEEEmembership{Senior~Member~IEEE},\\
Hongyang Du, 
Dusit Niyato,
\IEEEmembership{Fellow IEEE}
\thanks{Zonghang Li, Wenjiao Feng, Hongfang Yu, Long Luo, and Gang Sun are with the School of Information and Communication Engineering, University of Electronic Science and Technology of China, Chengdu 611731, China.}
\thanks{Weibo Cai is with the EZVIZ Network Co., Ltd., Hangzhou 310000, China.}
\thanks{Hongyang Du and Dusit Niyato are with the School of Computer Science and Engineering, Nanyang Technological University, 639798, Singapore.}
\thanks{Zonghang Li, Wenjiao Feng, and Weibo Cai are of equal contributions. The corresponding authors are Hongfang Yu (Email: \url{yuhfnetworklab@gmail.com}) and Long Luo (Email: \url{longluo.uestc@gmail.com}).}
\thanks{This work was supported in part by the National Natural Science Foundation of China (62102066), Sichuan Province Selected Funding for Postdoctoral Research Projects (TB2022032), and Open Research Projects of Zhejiang Lab (2022QA0AB02).}
}

\maketitle

\begin{abstract}
Distributed machine learning is becoming increasingly popular for geo-distributed data analytics, facilitating the collaborative analysis of data scattered across data centers in different regions. This paradigm eliminates the need for centralizing sensitive raw data in one location but faces the significant challenge of high parameter synchronization delays, which stems from the constraints of bandwidth-limited, heterogeneous, and fluctuating wide-area networks. Prior research has focused on optimizing the synchronization topology, evolving from starlike to tree-based structures. However, these solutions typically depend on regular tree structures and lack an adequate topology metric, resulting in limited improvements. This paper proposes \NAME, an adaptive and highly efficient communication scheduler designed to speed up parameter synchronization across geo-distributed data centers. First, it establishes an effective metric for optimizing a multi-root FAPT synchronization topology. Second, a network awareness module is developed to acquire network knowledge, aiding in topology decisions. Third, a multipath auxiliary transmission mechanism is introduced to enhance network awareness and facilitate multipath transmissions. Lastly, we design policy consistency protocols to guarantee seamless updates of transmission policies. Empirical results demonstrate that \NAME~significantly outperforms distributed training systems like MXNET, MLNET, and TSEngine, with a speedup of 6.5$\sim$9.2 times over MXNET.
\end{abstract}

\begin{IEEEkeywords}
Geo-distributed ML, Synchronization Topology, Communication Scheduling, Multipath Transmission.
\end{IEEEkeywords}
\section{Introduction}
\IEEEPARstart{A}{s} \new{data centers and big data become more geographically dispersed worldwide, transferring this vast amount of data over wide area networks (WANs) faces greater challenges, including slow transmission speeds and concerns about data privacy. These issues have led to the rise in popularity of geo-distributed big data analytics, which is supported by systems like MapReduce and Spark \cite{chen2018scheduling,niu2017multi}, as well as geo-distributed machine learning (GeoML) systems that are optimized for data parallelism \cite{zhou2019privacy,zhou2021tsengine} and model parallelism \cite{yuan2022decentralized}. Specifically, data centers use the data stored locally to train their machine learning (ML) model replicas, then synchronize the learned parameters or gradients with other centers to create an averaged model. This exchange process is referred to as parameter synchronization. However, unlike conventional distributed machine learning (DML) systems operating on large-scale, high-performance, homogeneous, and stable GPU clusters using Infiniband and RoCE technologies \cite{ren2017irdma,xue2019fast}, GeoML operates over WANs, which are known for high latency, low bandwidth, and a heterogeneous and dynamic nature. These constraints lead to significant delays in parameter synchronization.}

\new{Specifically, GeoML encounters three main challenges that lead to high parameter synchronization delays:
\begin{itemize}
    \item \textbf{Low Bandwidth.} WANs exhibit low bandwidth and high latency, especially between data centers in distant regions. However, traditional GeoML systems \cite{li2023placement,fan2023online,fan2023self,mi2020collaborative,cano2016towards,zhou2023nbsync,li2019online,lyu2019optimal} tend to use the inefficient starlike synchronization topology \cite{li2014scaling} between centers, further extending synchronization delays.
    \item \textbf{Network Heterogeneity.} Differences in transmission speeds between data centers are common \cite{hsieh2017gaia}, with slower centers blocking others during parameter synchronization. This blocking occurs not only at the parameter server \cite{li2020esync} but also at intermediate aggregation nodes \cite{zhou2021tsengine}.
    \item \textbf{Network Dynamics.} WAN resources are constantly fluctuating, with peak times causing congestion and off-peak times offering more bandwidth. Adapting to network changes is essential, as outdated communication strategies can lead to significantly reduced efficiency.
\end{itemize}}

\subsection{Prior Arts, Limitations and Motivations}
\new{Addressing these challenges requires optimizing the parameter synchronization topology. In response, current research \cite{geng2018hips,mai2015optimizing,wan2020rat,yang2023detfed,liu2021reconfigurable,sapio2021scaling,wang2020blink,zhang2021near,zhou2021tsengine,hsieh2017gaia,jeaugey2019massively,huang2021communication,reisizadeh2021codedreduce} has evolved from the starlike topology to more adaptable tree-based structures. These tree-based topologies link data centers in a hierarchical manner, aggregating model traffic at intermediate centers to lower transmission costs. However, studies like \cite{geng2018hips,huang2021communication,jeaugey2019massively,sapio2021scaling,mai2015optimizing,wan2020rat} depend on a known physical network topology that is regular, symmetric, and supports full-duplex bandwidth, and \cite{hsieh2017gaia,geng2018hips,mai2015optimizing,wan2020rat,liu2021reconfigurable,reisizadeh2021codedreduce} adopt static balanced trees for parameter synchronization, lacking adaptability to WAN heterogeneity and dynamics. Then, studies in \cite{liu2021reconfigurable,sapio2021scaling} require configuring programmable and optical switches, complicating deployment and potentially disrupting other communication services. Lastly, studies \cite{wang2020blink,zhang2021near,zhou2021tsengine} overlook the blockage delays at intermediate aggregation nodes, not accounting for the ``aggregate-forward'' nature of DML parameter synchronization, indicating a substantial opportunity for improvement.}

\new{These prior arts and limitations motivate us to consider the following questions:}

\new{\textit{(a) Which type of topology to optimize?} Given that physical WAN topologies are typically invisible and beyond the control of GeoML applications, optimizations based on physical topology is impractical. This motivates \textit{a shift toward optimizations based on the overlay network topology}, where data centers are considered as nodes and interconnected by VPN tunnels. In this way, \textit{the parameter synchronization topology can be optimized within this visible overlay network}.}

\new{\textit{(b) How to optimize the parameter synchronization topology?} Most current studies lack awareness of network resource availability, typically presuming that the underlying topology is symmetric and homogeneous, and result in the use of rigid balanced trees for parameter synchronization. Although some work \cite{wang2020blink,zhang2021near,zhou2021tsengine} has explored irregular tree-based topologies, these efforts overlook blockage delays in their optimization objectives. Motivated by these issues, \textit{network awareness and an appropriate objective that accounts for the ``aggregate-forward'' nature should be developed.}}

\subsection{Our Solution and Contributions}
This paper introduces \NAME, an adaptive and highly efficient communication scheduler optimized to speed up parameter synchronization across geo-distributed data centers, which are often constrained by scarce, heterogeneous, and dynamic network resources. To tackle the issue of an absent appropriate objective function, we design a \textit{topology metric} specifically for GeoML traffic with ``aggregate-forward'' pattern. This metric takes into account the transfer, blockage, and aggregation delays, focusing on non-leaf nodes that must await the arrival of model parameters from all child nodes before performing local aggregation and forwarding the aggregated data to their parent node, and yields a more precise performance evaluation. Based on this metric, we propose a \textit{multi-root fastest aggregation path tree (FAPT) topology} for efficient parameter synchronization, utilizing multiple root servers to distribute network load evenly and minimize total synchronization delay. The decision-making for this parameter synchronization topology benefits from the knowledge of network resource availability, facilitated by our \textit{passive network awareness module}. This module offers lightweight and precise network perception without additional probe traffic, enabling dynamic adjustments to the parameter synchronization topology to keep it up-to-date in fluctuating network conditions. Nevertheless, this passive approach may result in incomplete network knowledge and subpar topology decisions. To mitigate this, we introduce a \textit{multipath auxiliary transmission mechanism} that uses idle links outside the decision-making topology to assist the main path in model chunk transmission. In this way, idle links can be perceived and used for parallel transmission, further accelerating parameter transfer. Lastly, we implement a \textit{policy consistency protocol} to ensure smooth transitions between old and new topology configurations in response to network changes. Experiments on WANs emulated by Klonet \cite{ma2024klonet} showcase the substantial speedup of \NAME, achieving 6.5 to 9.2 times the training speed of the commonly used MXNET \cite{chen2015mxnet} with starlike topology and surpassing tree-based systems like MLNET \cite{mai2015optimizing} and TSEngine  \cite{zhou2021tsengine} in both dynamic and static network conditions.

The main contributions are summarized as follows.
\begin{itemize}
\item We designed a topology metric following the ``aggregate-forward'' pattern and proposed a multi-root FAPT topology for efficient parameter synchronization, achieving 7.6$\times$ speedup in heterogeneous and dynamic WANs.
\item We designed a passive network awareness module based on model parameter probes. It enables lightweight and precise link throughput measurement, resulting in a 20\% speedup without multipath auxiliary transmission.
\item We proposed multipath auxiliary transmission that utilizes idle links to offload a small portion of model transmission, enhancing network awareness and enabling parallel transmission, and achieves an additional 65\% speedup.
\item We implemented the prototype of \NAME~on top of MXNET, utilizing standardized PUSH/PULL interfaces to offer parameter synchronization services. We demonstrated its superior efficiency through comparative and ablation experiments, and we have made it open-source\footnote{The code is available at: \url{https://github.com/fengwenjiao/netstorm}.}.
\end{itemize}
\section{Preliminary Work and Motivations}\label{sec:related-work}
The synchronization topology plays an essential role in reducing synchronization delays and achieving efficient training. In this section, we review the evolution of synchronization topology in GeoML applications, tracing from starlike to tree-based topologies and from regular to irregular structures.

\begin{figure*}
\centering
\includegraphics[width=\textwidth]{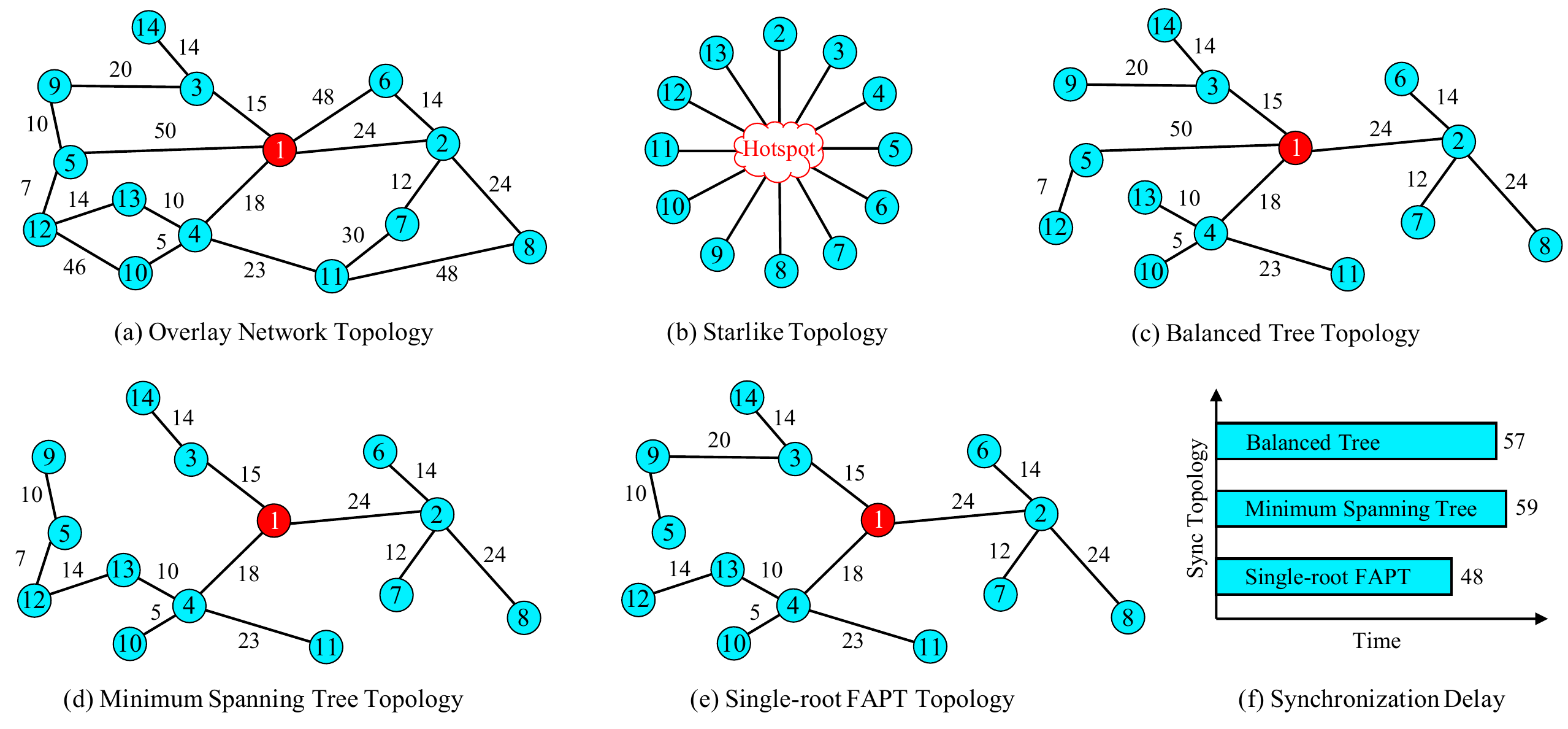}
\caption{Illustration of (a) the overlay network topology, synchronization topologies including (b) starlike structure and (c)-(e) three tree-based variants, and (f) the comparison of synchronization delays. The number on each edge indicates the delay incurred to transfer one unit of data through that link.}
\label{fig:tree-topos}
\end{figure*}

\subsection{Evolving from Starlike to Tree-based Topologies}
\new{Parameter Server (PS) \cite{li2014scaling} is a popular synchronization framework extensively used by DML systems like TensorFlow \cite{abadi2016tensorflow}, MXNET \cite{chen2015mxnet}, and NBSync \cite{zhou2023nbsync}, as well as by GeoML systems \cite{li2023placement,fan2023online,fan2023self,li2019online,lyu2019optimal}. This framework features a starlike synchronization topology with a central parameter server and multiple workers, as illustrated in Figure \ref{fig:tree-topos}b. In each training round, workers train their local model replicas, push the learned parameters or gradients to the parameter server, and pull the updated global model parameters to finish off this round. To improve communication efficiency in GeoML training, Li \textit{et al.} \cite{li2023placement} propose selecting a communication-cost-effective node as the parameter server. Fan \textit{et al.} \cite{fan2023online} address bandwidth contention across multiple training jobs by employing online traffic scheduling. Furthermore, Fan \textit{et al.} \cite{fan2023self} introduce an adaptive gradient quantization method, enabling workers to adjust quantization based on their bandwidth, thus lowering communication costs and alleviating delays caused by stragglers. Nonetheless, these systems exhibit inherent limitations: all workers should establish direct connections to the parameter server, leading to significant congestion at this node and posing challenges to system scalability.}

\new{Consequently, research has shifted toward tree-based synchronization topologies, aiming for higher efficiency and scalability. These topologies involve a node as the root, with other nodes as leaves or intermediaries, forming a tree structure. Model parameters flow from the leaves, aggregate through intermediate nodes, and ultimately converge at the root node. Notably, non-leaf nodes must aggregate model parameters from both their child nodes and themselves. Tree-based topologies offer notable benefits, such as reduced model traffic on the network, alleviated congestion at the parameter server, and better utilization of idle links between non-root nodes. Indeed, studies \cite{huang2021communication,zhang2021near} have confirm that spanning trees are the best choice for parameter synchronization.}

\new{However, due to the lack of knowledge about network heterogeneity, most systems have defaulted to using balanced tree structures, as illustrated in Figure \ref{fig:tree-topos}c, presuming homogeneous network resources. Within these, balanced binary trees, as discussed in \cite{liu2021reconfigurable, jeaugey2019massively}, are characterized by having at most two child nodes per non-leaf node and a depth variation of no more than one, offering logarithmic communication complexity and greater efficiency than flat architectures like PS. Moreover, balanced $k$-way trees, as investigated in \cite{mi2020collaborative,cano2016towards,mai2015optimizing,wan2020rat,luo2020plink,geng2018hips,reisizadeh2021codedreduce}, reduce transmission hops through more child nodes, further enhancing communication efficiency. Notably, even though \cite{mi2020collaborative,cano2016towards} target GeoML applications, they still use the flat PS architecture for inter-data center synchronization.}

\subsection{Evolving from Regular to Irregular Topologies}
\new{Recent studies \cite{wang2020blink,zhang2021near,zhou2021tsengine} highlight that regular tree structures fall short in addressing the resource heterogeneity in WANs, as they often lead to bottleneck links and synchronization blockages. Hence, \textit{incorporating network awareness to develop irregular tree-based topologies becomes essential to minimize synchronization delays.} Zhou \textit{et al.} \cite{zhou2021tsengine} employ a minimum spanning tree to build their TSEngine system, as illustrated in Figure \ref{fig:tree-topos}d, prioritizing transmission over links with the highest data throughput to avoid bottleneck connections. Wang \textit{et al.} \cite{wang2020blink} use packing spanning trees to maximize flow from a selected root to all other nodes. Zhang \textit{et al.} \cite{zhang2021near} recognize that in real-world settings, only some nodes might be active, and propose using Steiner forest for effective parameter aggregation and broadcasting. In this model, each Steiner tree, rooted at a different node, handles the transmission of different model segments, aiming to maximize network bandwidth utilization and minimize the aggregation and broadcast times dependent on the minimum link bandwidth in the tree. Notably, when all nodes are active, Steiner trees function as spanning trees.}

\subsection{``Aggregate-Forward'' Needs New Topology Metric}
\new{Existing works like \cite{wang2020blink} focusing on maximizing data flow, or \cite{zhang2021near,zhou2021tsengine} aiming to improve bottleneck link bandwidth, view DML traffic as regular data streams, overlooking its distinctive ``aggregate-forward'' nature. In typical routing optimization, data packets travel from the source through the network via multiple hops to reach their destination. While this process superficially resembles the ``aggregate-forward'' mode, the difference lies in the network devices' behavior: they merely forward the packets based on routing tables, without performing any data aggregation.}

\new{Optimizing the parameter synchronization topology is distinct from conventional routing optimization, as DML traffic requires packet aggregation at non-leaf nodes (not on network devices). In this process, non-leaf nodes must collect data packets from their child nodes, aggregate them locally, and then forward the result to the parent node. This transmission mode, termed ``aggregate-forward'', forces non-leaf nodes to wait until they have received all packets from their child nodes, resulting in \textit{blockage delays}. Consequently, as illustrated in Figures \ref{fig:tree-topos}e and \ref{fig:tree-topos}f, it's possible to identify a more efficient tree structure with a lower synchronization delay.}

\new{The ``aggregate-forward'' nature of DML traffic goes beyond basic streaming data transfer, involving hop-by-hop transmission across the synchronization topology, with blockage delays accumulating at non-leaf nodes. This motivates a \textit{reevaluation of existing topology metrics and optimization objectives} to account for these delays and prompts the \textit{redesign of the communication scheduler} for GeoML systems.}
\section{Problem Modeling and Challenges}
\label{sec:problem-modeling}

\subsection{Definitions and Problem Model}\label{sec:definitions-and-problem-model}
To identify a more efficient synchronization topology, we should first define the synchronization delay for a given topology, and then pinpoint the topology that minimizes this delay. The new metrics and objectives should account not only for hop-by-hop transmission delays but also for blockage delays and processing delays.

\vspace{1.5mm}
\textit{(a) How to quantify the synchronization delay?}
\vspace{1.5mm}

Assuming we have knowledge of the overlay network topology connecting geo-distributed data centers and have measured the throughput of their links, this information can be used to identify an effective synchronization topology for GeoML parameter exchange. As the first step, we define the total delay on a given path as the sum of transfer delays, blockage delays, and processing delays along that path.

\begin{definition}[Delay of a Path]
\label{def:path-delay}
For a transmission path $p_l=\{v_1,v_2,\cdots,v_n\}$, where $v_1$ is the leaf and $v_n$ the root, with each $v_i$ representing the $i$-th node in the path. Let $e$ be any link on this path, $w_{\mathrm{trans}}(e)$ denote the transfer delay for $e$, $w_{\mathrm{block}}(v_i)$ the blockage delay at node $v_i$ due to waiting for its slower child nodes, and $w_{\mathrm{proc}}(v_i)$ the processing delay for node $v_i$ to perform aggregation computations. Then, the synchronization delay $w(p_l)$ for path $p_l$ can be defined as:
\begin{equation} 
w(p_l)=\underset{\mathrm{transfer}\ \mathrm{delays}}{\underbrace{\sum_{e\in p_l}{w_{\mathrm{trans}}}(e)}}+\underset{\mathrm{blockage}\ \mathrm{delays}\ \mathrm{and}\ \mathrm{processing}\ \mathrm{delays}}{\underbrace{\sum_{v_i\in p_l\backslash \{v_1\}}{\left( w_{\mathrm{block}}\left( v_i \right) +w_{\mathrm{proc}}\left( v_i \right) \right)}}}
\end{equation}
\end{definition}

In our implementation, we overlap aggregation computations with parameter transmission, allowing us to neglect processing delays because they are negligible compared to the significant transfer delays, as shown in Figure \ref{fig:time-overlap}. This focuses our efforts on minimizing transfer delays and blockage delays, then we can obtain the following theorem.

\begin{theorem}[Synchronization Delay of a Tree-based Topology]
\label{theorem:tree-completion-time}
For a tree topology $T_{v_i}$ with root $v_i$, where each link $e$ in the edge set $E$ carries a positive weight $w_{\mathrm{trans}}(e)$ denoting its transfer delay. The synchronization delay for this tree, $w(T_{v_i})$, is defined as the maximum sum of transfer delays $w_{\mathrm{trans}}(e),\forall e\in p_l$ along any path $p_l$ from the leaves to the root $v_i$ within the tree $T_{v_i}$, represented as:
\begin{equation}
\label{eq:tree-completion-time}
w(T_{v_i}) = \max_{p_l \in \mathcal{P}(T_{v_i})} \sum_{e \in p_l} w_{\mathrm{trans}}(e)
\end{equation}
where $p_l$ represents a possible path within all feasible path set $\mathcal{P}(T_{v_i}))$ extending from any leaf to the root $v_i$.
\end{theorem}

With Theorem \ref{theorem:tree-completion-time} established, we revisit how the synchronization delay in Figure \ref{fig:tree-topos}f is calculated. Taking the balanced tree topology shown in Figure \ref{fig:tree-topos}c as an example, subtree $T_{v_2}$ aggregates model parameters from four nodes $\{v_2,v_6,v_7,v_8\}$. As $T_{v_2}$ is an aggregation tree, the parent node $v_2$ must wait until it has received all the model parameters from its child nodes to perform the aggregation, taking a time of $w(T_{v_2})=\max \left\{ w_{\mathrm{trans}}\left( e_{6,2} \right) ,w_{\mathrm{trans}}\left( e_{7,2} \right) ,w_{\mathrm{trans}}\left( e_{8,2} \right) \right\} =24$ time units, where $e_{i,j}$ represents the link from node $i$ to node $j$. Similarly, the synchronization delays for the subtrees $T_{v_3}=\{v_3,v_9,v_{14}\},T_{v_4}=\{ v_4,v_{10},v_{11},v_{13}\},T_{v_5}=\{v_5,v_{12}\}$ are computed, taking $w\left( T_{v_3} \right) =20,w\left( T_{v_4} \right) =23,w\left( T_{v_5} \right) =7$ time units, respectively. For the whole tree $T_{v_1}$, with the root node $v_1$ being the parent of subtrees $\{v_{2},v_{3},v_{4},v_{5}\}$, it requires a total of
\begin{align}\nonumber
w\left( T_{v_1} \right) &=\max \left\{ \begin{array}{c}
	w\left( T_{v_2} \right) +w_{\mathrm{trans}}\left( e_{2,1} \right) ,w\left( T_{v_3} \right) +w_{\mathrm{trans}}\left( e_{3,1} \right)\\
	w\left( T_{v_4} \right) +w_{\mathrm{trans}}\left( e_{4,1} \right) ,w\left( T_{v_5} \right) +w_{\mathrm{trans}}\left( e_{5,1} \right)\\
\end{array} \right\} 
\\ \nonumber
&=\max \{48,35,41,57\} =57
\end{align}
time units to complete parameter synchronization. The same method applies to the calculations in Figures \ref{fig:tree-topos}d and \ref{fig:tree-topos}e.

Theorem \ref{theorem:tree-completion-time} offers a new approach to optimizing the synchronization topology. Based on this theorem, the objective is to find the path with the minimal cumulative transfer delay, achievable through shortest path algorithms. The resulting topology, illustrated in Figure \ref{fig:tree-topos}e, exhibits the lowest synchronization delay relative to other tree structures. Therefore, it is referred to as the ``Fastest Aggregation Path Tree''.

\begin{definition}[Fastest Aggregation Path Tree]
\label{def:fapt}
The Fastest Aggregation Path Tree (FAPT) is a parameter synchronization topology structured as a tree, characterized by its minimal synchronization delay. It is formally defined as:
\begin{equation}
T_{v_i}=\mathrm{arg}\underset{T\in \mathcal{T}_{v_i}}{\min}\max_{p_l\in \mathcal{P}\left( T \right)} \sum_{e\in p_l}{w}_{\mathrm{trans}}(e)
\end{equation}
where $\mathcal{T}_{v_i}$ is the set of all feasible trees rooted at $v_i$, and $\mathcal{P}(T)$ is the set of paths in tree $T$ from all leaves to the root.
\end{definition}

\begin{figure*}
\centering
\includegraphics[width=\textwidth]{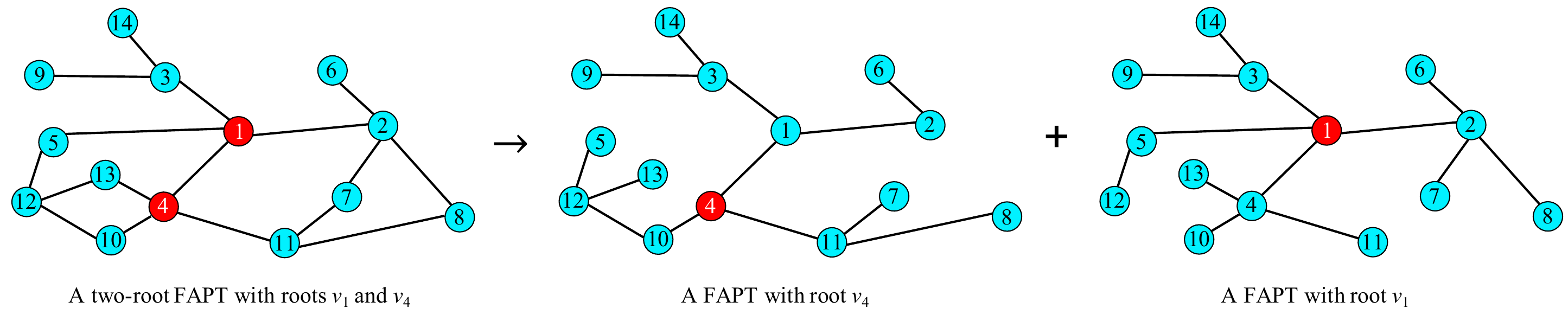}
\caption{An illustration of a two-root FAPT with the root set $R=\{v_1,v_4\}$.}
\label{fig:union-tree-cover}
\end{figure*}

\vspace{1.5mm}
\textit{(b) How to define the objective of topology optimization?}
\vspace{1.5mm}

While FAPT offers superior synchronization efficiency relative to other tree structures, its native implementation uses only one parameter server. In contrast, within the PS architecture, it is common to have multiple parameter servers, each handling different segments of model parameters to distribute storage and traffic loads. To further reduce synchronization delay, FAPT should be extended to support multiple roots, with each root leading a separate tree. Consequently, the optimization problem transforms into a min-max problem for multi-root FAPT, aiming to minimize the maximum synchronization delay across these trees.

\begin{definition}[Multi-Root Fastest Aggregation Path Tree]
\label{def:r-rooted-union-tree-cover}
Let $G=(V,E)$ be an undirected graph with positive edge weights $w:E\rightarrow\mathbb{R}^+$, and $R\subset V$ be a set of root nodes. A multi-root FAPT, denoted as $\bar{G}_R$, is a subgraph formed by the union of 
$N$ multiple FAPTs, $\mathcal{T}(\bar{G}_R)=\{T_{v_1},\cdots,T_{v_N}\}$, where $N=|R|$, $\bar{G}_R=\bigcup_{i=1}^{N}{T_{v_i}}\subset G$, and each tree $T_{v_i}$ has a unique root node $v_i\in R$. The trees in $\mathcal{T}(\bar{G}_R)$ can share links. The cost of a multi-root FAPT is defined as the maximum synchronization delay among all the trees in $\mathcal{T}(\bar{G}_R)$, i.e., $J=\max_{T_{v_i}\in\mathcal{T}(\bar{G}_R)}w(T_{v_i})$.
\end{definition}

Figure \ref{fig:union-tree-cover} shows an example of a two-root FAPT. It includes two roots, $v_1$ and $v_4$, each leading a separate FAPT. Intuitively, the two-root FAPT is the union of these two FAPTs, with the root set $R=\{v_1,v_4\}$.

With the definition of multi-root FAPT, our objective is to find a multi-root FAPT topology $\bar{G}_R$ that minimizes the cost $J$. The formal problem definition is as follows:
\begin{align}
\label{eq:min-max-r-rooted-tree-cover-obj}
\min_{\bar{G}_R \subset G}\max_{T_{v_i}\in\mathcal{T}(\bar{G}_R)} \quad & w(T_{v_i}) \\
\text{s.t.}\quad &\bar{G}_R = \bigcup_{i=1}^{N}{T_{v_i}},\label{eq:cons-g} \\
&|V(T_{v_i})| = |V|,~\forall T_{v_i} \in \mathcal{T},\label{eq:cons-v} \\
&w(T_{v_i})=\max_{p_l\in\mathcal{P}}\sum_{e\in p_l}w_{\mathrm{trans}}(e),\label{eq:cons-w} \\
\label{eq:cons-0}
&w_{\mathrm{trans}}(e) > 0,~\forall e\in E.
\end{align}
\new{The first constraint Eq. \eqref{eq:cons-g} ensures that the resulting multi-root FAPT is the union of $N$ FAPTs $T_{v_i}$ corresponding to the $N$ roots in the set $R$. The second constraint Eq. \eqref{eq:cons-v} guarantees that each tree $T_{v_i}$ covers all nodes in $V$. The third constraint Eq. \eqref{eq:cons-w} defines the synchronization delay for tree $T_{v_i}$. The last constraint Eq. \eqref{eq:cons-0} ensures that the edges have positive weights.}

\subsection{Main Challenges}\label{sec:problem-model-challenges}
To solve the multi-root FAPT problem and develop an adaptive distributed training system, four challenges should be addressed:

\textit{(a) Complexity in solving the multi-root FAPT problem.} 
The problem model involves selecting $N$ distinct nodes from $V$ to establish the root set $R$ and then determining a multi-root FAPT topology based on $R$. The solution space is enormous, with $\mathrm{C}_{|V|}^{N}$ potential combinations, where the complexity increases exponentially as the number of nodes $|V|$ grows. However, given the dynamic nature of WAN conditions, there is a need for frequent resolutions of this problem to timely update the synchronization topology. Therefore, the solution must be computational efficient, making the problem challenging.

\textit{(b) Need for lightweight and real-time network awareness.} 
Understanding network resource availability is essential for guiding the decision-making and adjustment of the synchronization topology. The primary metric in this context is link throughput, typically measured through active probing methods that introduce extra traffic to the network. However, these methods are not good choices for real-time measurements as the probe traffic competes with GeoML traffic and burdens the network. Therefore, a passive probing approach is preferable, using the GeoML traffic itself for probing.

\textit{(c) Avalanche effect in passive network awareness.}
The decision-making for synchronization topology depends on link throughput data provided by passive network awareness. For optimal topology decision, full network knowledge is essential. However, passive network awareness is constrained as it measures links only within the determined topology, leaving external links unmeasured. This results in partial network knowledge and suboptimal topology decisions. Furthermore, network changes might not be timely detected, leading to outdated topology. Hence, an auxiliary mechanism is needed to enhance network awareness, ensuring that all the links can be utilized and measured.

\textit{(d) Inconsistent topology leads to confusion.} As the synchronization topology can update, inconsistencies may arise across local topologies at different nodes, with old and new configurations coexisting, potentially causing deadlocks and packet loss. Therefore, it is essential to design a consistency protocol that guarantees correct synchronization across distributed nodes.

For challenge (a), Section \ref{sec:multi-root-fapt-topo} introduces a fast algorithm to solve the multi-root FAPT problem. To address challenge (b), Section \ref{sec:network-awareness} implements a passive network awareness module based on model parameter probes. In response to challenge (c), Section \ref{sec:multipath-aux-transmission} proposes a multipath auxiliary transmission mechanism. Finally, for challenge (d), Section \ref{sec:policy-consistency} elaborates on a consistency protocol designed to ensure a smooth transition to new configurations during updates.
\section{Multi-root FAPT Topology}\label{sec:multi-root-fapt-topo}
In this section, we elaborate on the main idea and algorithm design to construct the multi-root FAPT topology and provide details on our implementation.

\subsection{Main Idea}\label{sec:multi-root-fapt-topo-main-idea}
To solve the problem outlined in Eqs. \eqref{eq:min-max-r-rooted-tree-cover-obj}$\sim$\eqref{eq:cons-0}, we first consider a simplified case where $N=1$. In this case, the synchronization topology involves only one root, $v_i$, yielding $\mathrm{C}_{|V|}^{1}=|V|$ possible choices for $v_i$. In this context, the multi-root FAPT reduces to a single FAPT. We can formally define this simplified problem as follows:
\begin{align}
\min_{T_{v_i} \subset G} \max_{p_l\in\mathcal{P}(T_{v_i})} \quad & \sum_{e\in p_l}w_{\mathrm{trans}}(e)\label{eq:min-max-new-obj} \\
\text{s.t.}\quad & \forall v_i\in V,\\
\nonumber & \text{Eqs. }\eqref{eq:cons-v}, \eqref{eq:cons-0}.
\end{align}

This problem focuses on identifying a root $v_i$ from the set $V$ and constructing a FAPT $T_{v_i}$ that exhibits the minimal synchronization delay. According to Theorem \ref{theorem:tree-completion-time}, the synchronization delay is determined by the maximum sum of transfer delays along any path from the leaves to the root. As analyzed in Section \ref{sec:definitions-and-problem-model}, this problem can be resolved using the shortest path algorithm, as minimizing the cumulative transfer delay for all paths inherently minimizes the delay on the slowest path. Thus, for each candidate root $v_i\in V$, the shortest path algorithm can identify its FAPT, making it easy to pinpoint the root with minimal synchronization delay.

In the more general case where $N>1$, we simplify the analysis by predefining the root set $R$. Then, we can construct a tree set $\bar{\mathcal{T}}_{v_i}$ for each root $v_i\in R$, containing all feasible trees rooted at $v_i$. The problem then shifts from constructing a subgraph $\bar{G}_R$ from the full graph $G$ to identifying the optimal FAPT $T_{v_i}$ from its set $\bar{\mathcal{T}}_{v_i}$, as follows:
\begin{align}
\min_{T_{v_i} \in \bar{\mathcal{T}}_{v_i}}\max_{T_{v_i}\in\mathcal{T}^{\prime}} \quad & w(T_{v_i}) \\
\text{s.t.}\quad &\bar{\mathcal{T}}_{v_i}=\{T_{v_i},\forall T_{v_i}\subset G~\mathrm{with~root}~v_i\},\label{eq:cons-t}\\
\nonumber &\text{Eqs. }\eqref{eq:cons-v}, \eqref{eq:cons-w}, \eqref{eq:cons-0}.
\end{align}
The objective is to find a set of trees $\mathcal{T}^{\prime}=\{T_{v_i},~\forall v_i\in R\}$ from the feasible set $\bar{\mathcal{T}}_{v_i}$, such that the maximum synchronization delay among $\mathcal{T}^{\prime}$ is minimized. 

To achieve this goal, minimizing the synchronization delay for all trees in $\mathcal{T}^{\prime}$ is viable, given the independence of any two trees in the set. This independence holds as trees sharing the same link occupy distinct portions of the bandwidth to synchronize different segments of model parameters, as shown in Figure \ref{fig:multi-server-load-balance}, thereby preventing resource contention. Consequently, minimizing the synchronization delay for all trees inherently reduces the delay for the slowest tree. Hence, the problem can be reformulated as:
\begin{align}
\min_{T_{v_i} \in \bar{\mathcal{T}}_{v_i}} \quad & \sum_{T_{v_i}\in\mathcal{T}^{\prime}} w(T_{v_i})\label{eq:min-new-obj} \\
\nonumber\text{s.t.}\quad &\text{Eqs. }\eqref{eq:cons-v}, \eqref{eq:cons-w}, \eqref{eq:cons-0}, \eqref{eq:cons-t}.
\end{align}
By applying Eq. \eqref{eq:cons-w} to Eq. \eqref{eq:min-new-obj}, it becomes clear that this problem has a similar structure to Eq. \eqref{eq:min-max-new-obj}. Thus, it can be solved by identifying the FAPT for each root $v_i\in R$.

Finally, we return to the original problem of identifying the solution without a predefined root set $R$. To identify $R$, a straightforward yet effective approach can be employed. Given that the synchronization delay $w(T_{v_i})$ for each FAPT $T_{v_i}$, rooted at $v_i$ for all $v_i\in V$, has been calculated as per Definition \ref{theorem:tree-completion-time}, we can define a quality score $q_i=\frac{1}{w(T_{v_i})}$. This score represents the synchronization efficiency of the FAPT $T_{v_i}$, with higher values of $q_i$ indicating better performance. Then, the root set $R$ can be established by selecting the nodes with the highest $N$ quality scores.

\subsection{Algorithm Design}
This section presents algorithms to identify all the fastest aggregation paths (via Algorithm \ref{alg:find-all-fapt-trees}) and construct the multi-root FAPT topology (via Algorithm \ref{alg:r-rooted-fapt-topo-construct}), using the outcomes from Algorithm \ref{alg:find-all-fapt-trees}.

\begin{algorithm}[t]
\caption{\textsc{Find-Fastest-Aggregation-Paths}}
\label{alg:find-all-fapt-trees}
\begin{algorithmic}[1]
\small
\Require An undirected graph $G=(V,E)$; the transfer delay $w_{\mathrm{trans}}(e)$ for each link $e$; a set of root $R$; number of roots $N$.
\Ensure The set of root $R$; the set of node sequences for all fastest aggregation paths $H$.
\If{the root set is not determined before, i.e., $R=\varnothing$}
\State{Calculate the quality score $q_i$ for each node $v_i\in V$;}
\State{Sort all nodes by quality score;}
\State{Select the top $N$ nodes with the highest quality scores to form the root set $R$;}
\EndIf
\For{each node $v_i\in V$}
\State{Use shortest path algorithm to calculate the node sequence $p_{i\rightarrow j}$ from $v_i$ to every other $v_j$;}
\State{Add the sequence to set: $H=H\cup \left\{ p_{i\rightarrow j},\forall v_j\in V\backslash v_i \right\} $;}
\EndFor
\State{\Return{$R, H$};}
\end{algorithmic}
\end{algorithm}

\vspace{1.5mm}
\textit{(a) Identify root set and find all fastest aggregation paths.}
\vspace{1.5mm}

Algorithm \ref{alg:find-all-fapt-trees} is designed to identify the root set $R$ and find the fastest aggregation path between any pair of nodes. Initially, to form $R$, the algorithm iterates over all nodes to calculate the quality scores for FAPTs with each node as the root. It then selects the top $N$ nodes with the highest scores to initialize $R$. This root set $R$ is only set up once during the first run of Algorithm \ref{alg:find-all-fapt-trees} and remains fixed thereafter. This design is to avoid frequent model parameter migrations across WANs, as changing these roots would require transferring model parameter segments from old to new roots, leading to considerable overhead.

The next step is to search for the set of fastest aggregation paths $H$ between all node pairs. Following the analysis in Section \ref{sec:multi-root-fapt-topo-main-idea}, for each node pair $(v_i, v_j)$, the shortest path algorithm is applied to find the fastest aggregation path from $v_i$ to $v_j$, denoted as $p_{i\rightarrow j}$. If $v_i$ is unreachable from $v_j$, $p_{i\rightarrow j}$ yields an empty set. All these paths are compiled into a set $H$. This path set $H$, along with the root set $R$, is subsequently used in Algorithm \ref{alg:r-rooted-fapt-topo-construct} to construct the multi-root FAPT topology.

\vspace{1.5mm}
\textit{(b) Main: Construct the multi-root FAPT topology.}
\vspace{1.5mm}

\begin{algorithm}[t]
\caption{\textsc{Build-Multi-Root-FAPT-Topology}}
\label{alg:r-rooted-fapt-topo-construct}
\begin{algorithmic}[1]
\small
\Require An undirected graph $G=(V,E)$; the data throughput $s(e)$ for each link $e$; a set of root $R$; number of roots $N$.
\Ensure A multi-root FAPT topology $\bar{G}_R$ rooted at set $R$.
\State{Reset the transfer delay $w_{\mathrm{trans}}(e)$ for each link $e\in E$ using the latest link throughput $s(e)$: $w_{\mathrm{trans}}(e) = \frac{1}{s(e)}, \forall e \in E$;}
\State{$R,H\gets\textsc{Find-Fastest-Aggregation-Paths}(G,w_{\mathrm{trans}},R,N)$;}
\For{each root $v_i\in R$}
\State{Initialize an empty tree $T_{v_i}$ with only root node $v_i$;}
\For{each non-root node $v_j\in V\backslash R$}
\State{Pop a fastest aggregation path $p_{i\rightarrow j}$ from the set $H$;}
\State{Traverse $p_{i\rightarrow j}$ to establish parent-child relationships between adjacent nodes along the path;}
\EndFor
\EndFor
\State{\Return{\( \bar{G}_R = \bigcup_{v_i \in R} T_{v_i} \);}}
\end{algorithmic}
\end{algorithm}

Algorithm \ref{alg:r-rooted-fapt-topo-construct} serves as the \textit{main function} for topology decision, \new{executed by a central scheduler} every \texttt{UPDATE\_TIME} seconds, to refresh the synchronization topology in response to network dynamics. Each time it runs, the algorithm updates the transfer delay for each link based on the most recent link throughput measurement, thus basing decisions on current network conditions. Then, it invokes Algorithm \ref{alg:find-all-fapt-trees} to identify or update the root set $R$ and the fastest aggregation path set $H$. Utilizing these, we can construct FAPTs for each root and combine them to create a multi-root FAPT topology $\bar{G}_R$. The paths within $\bar{G}_R$ are primary paths for parameter synchronization. Notably, the root set $R$ is empty during the first run of Algorithm \ref{alg:r-rooted-fapt-topo-construct} and is initialized via Algorithm \ref{alg:find-all-fapt-trees}, after which it is fixed in subsequent calls.

\textit{Complexity Analysis.}
The time complexity of Algorithm \ref{alg:r-rooted-fapt-topo-construct} comes from two aspects: identifying the fastest aggregation paths and constructing the multi-root FAPT topology. Algorithm \ref{alg:find-all-fapt-trees} incurs a complexity of $\mathcal{O}(|V|^3)$. In Algorithm \ref{alg:r-rooted-fapt-topo-construct}, the tasks include updating the transfer delay for each link, calling Algorithm \ref{alg:find-all-fapt-trees}, and creating a FAPT for each root in the set $R$, leading to overall time complexity of $\mathcal{O}((N+|V|)|V|^2-N^2|V|+|E|)$. In general, $N\le|V|<|E|$.

The low runtime latency of Algorithm \ref{alg:r-rooted-fapt-topo-construct}, which needs frequent execution, is advantageous as it allows the system to quickly adapt to network changes and update the synchronization topology in a timely manner.

\subsection{Implementation Details}\label{sec:multi-root-fapt-alg-implementation}

\vspace{1.5mm}
\textit{(a) Multi-Root Load Balacing}
\vspace{1.5mm}

\begin{figure}[t]
\centering
\includegraphics[width=.4\textwidth]{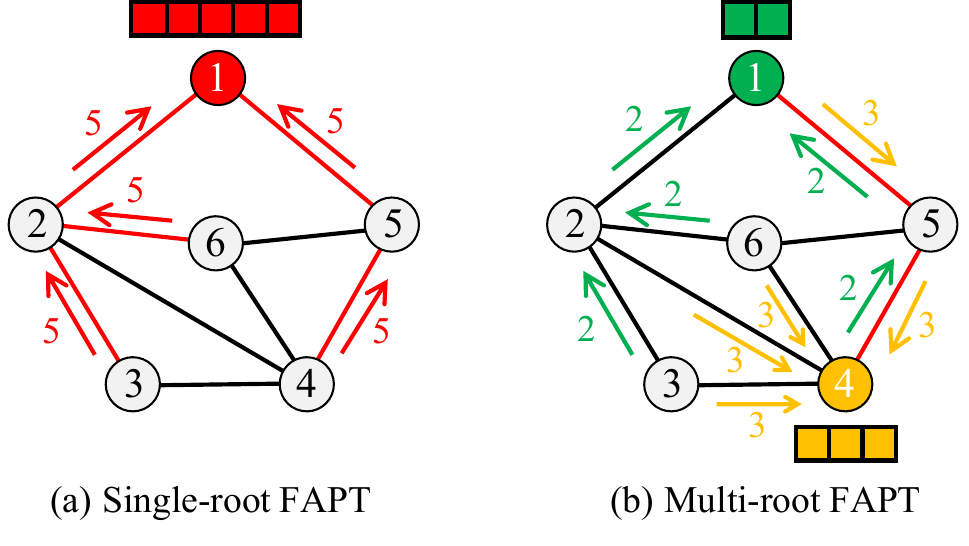}
\caption{An example of single-root and multi-root FAPT. Red edges indicate hotspot links. Numbers on edges indicate the number of chunks in transmit. Red, green, and yellow nodes are the roots, which manage a different number of chunks according to their availability.}
\label{fig:multi-server-load-balance}
\end{figure}

The multi-root design in Definition \ref{def:r-rooted-union-tree-cover} involves splitting the raw parameter tensor into chunks and allocating these chunks among multiple roots. In our implementation, we define \texttt{CHUNK\_SIZE} as the maximum size for these chunks. Large tensors are divided into smaller chunks of size \texttt{CHUNK\_SIZE}, while tensors smaller than \texttt{CHUNK\_SIZE} remain unchanged. The allocation of these chunks to the roots is guided by the quality scores, with each root receiving a proportion of chunks determined by $\frac{q_{i}}{\sum_{v_j\in R}{q_{j}}}$, where $q_{i}$ is the quality score. In brief, roots with higher scores have greater synchronization efficiency and manage more chunks. Figure \ref{fig:multi-server-load-balance} illustrates the difference between single-root and multi-root setups. In the single-root setup, one node manages all model parameters, leading to network hotspots and inefficient use of network links. Conversely, in the multi-root approach, each root manages a portion of the model parameters, which helps disperse network traffic, reduce network hotspots, improve the utilization of idle links, and also enhance network awareness.

\vspace{1.5mm}
\textit{(b) Time Overlap Between Transmission and Aggregation}
\vspace{1.5mm}

\begin{figure}[t]
\centering
\includegraphics[width=.5\textwidth]{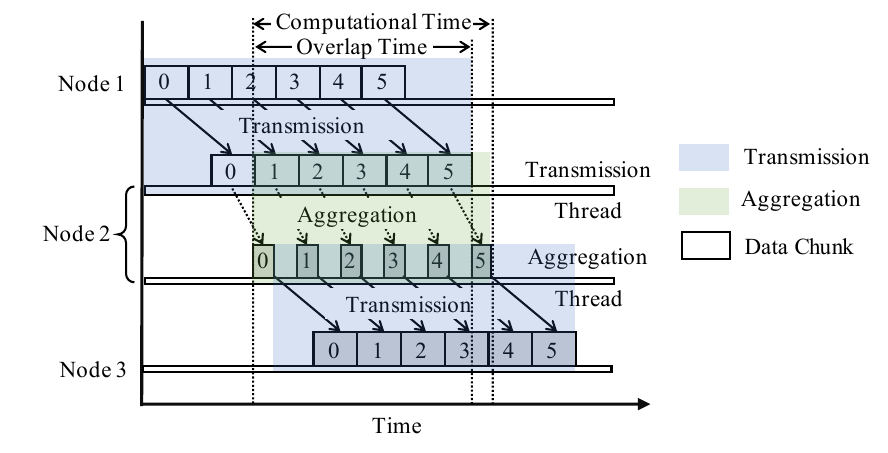}
\caption{The time overlap between transmission and aggregation.}
\label{fig:time-overlap}
\end{figure}

In Section \ref{sec:definitions-and-problem-model}, we assumed that the processing delay for aggregation is negligible. This assumption allows us to focus on optimizing the transfer and blockage delay, leading to Theorem \ref{theorem:tree-completion-time}. This assumption is reasonable, because there are time overlaps between aggregation and transmission between adjacent nodes, as illustrated in Figure \ref{fig:time-overlap}. As previously mentioned, parameter tensors are split into several similarly sized small chunks, whose aggregation and transmission can happen independently and concurrently. For example, while chunk 1 is in transmission, chunk 0 is being aggregated. Likewise, the transmission of chunks 1 to 5 and the aggregation of chunks 0 to 4 can be executed concurrently. This time overlap suggests that only the aggregation time for the last chunk is of concern. Given that this last chunk is constrained to be smaller than \texttt{CHUNK\_SIZE}, its aggregation time is so short that it can be disregarded compared to the transfer delay of the full tensor.

\section{Passive Network Awareness with Native Probes}\label{sec:network-awareness}
To address challenge (b) discussed in Section \ref{sec:problem-model-challenges}, this section introduces a passive network awareness module that uses native probes, that is, the parameter synchronization traffic generated by the distributed training system itself. By measuring the transfer delay of model chunks across links, this module can estimate the link throughput without injecting extra probing traffic. Thus, it offers efficient and unobtrusive measurements, ensuring a minimal impact on the network.

\begin{figure}[t]
\centering
\includegraphics[width=0.48\textwidth]{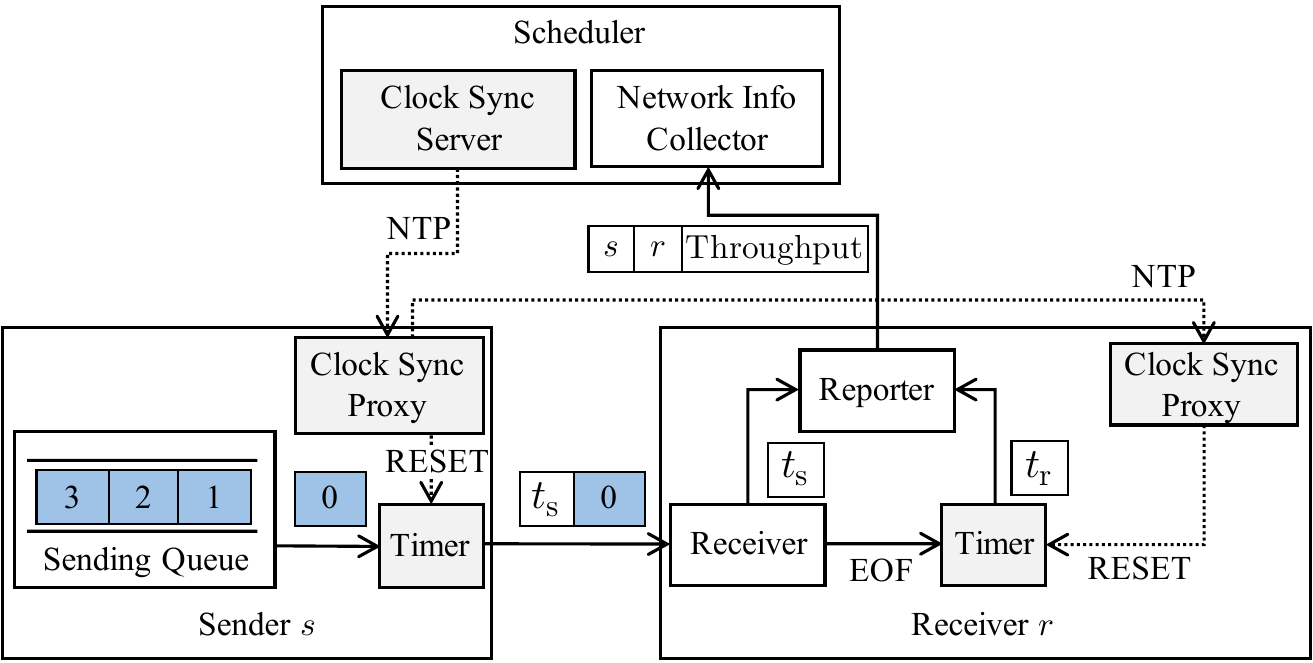}
\caption{Implementation of our network awareness module.}
\label{fig:awareness-module}
\end{figure}

The architecture of our network awareness module is shown in Figure \ref{fig:awareness-module}. It is implemented on both the scheduler and the worker nodes. On the scheduler side, it collects network reports from the worker nodes and manages clock synchronization. These network reports feed into Algorithm \ref{alg:r-rooted-fapt-topo-construct} to facilitate decision-making or updates to the synchronization topology. On the worker node side, every worker node features a timer to time the transfer delay, a reporter to estimate the link throughput and report it to the scheduler, and a proxy for clock synchronization.

In Figure \ref{fig:awareness-module}, parameter tensors have been divided into several similarly sized small chunks for transmission. As the sender $s$ begins data transmission, it picks a model chunk $i$ of size $S^i$ from the sending queue and records the sending timestamp $t_{\mathrm{s}}^i$. This timestamp will be attached to the model chunk message and sent to the receiver. Upon arrival, the receiver $r$ records the arrival timestamp $t_{\mathrm{r}}^i$. By repeating this for each chunk, a set of triplets $\{(t_{\mathrm{s}}^i, t_{\mathrm{r}}^i, S^i)| i=1,\cdots,I\}$ is collected, with $I$ denoting the number of chunks, specified as \texttt{PROBE\_CHUNK\_NUM}, used to estimate the average link throughput $\tau$. This throughput is calculated as follows:
\begin{equation}
\label{eq:throughput}
\tau=\sum_{i=1}^I\frac{S^i}{(t_{\mathrm{r}}^i-t_{\mathrm{s}}^i)I}
\end{equation}
The computed throughput $\tau$, along with the identifiers of the sender $s$ and receiver $r$, is then reported to the scheduler for network awareness.

\begin{proposition}\label{prop:measure-accuracy}
The link throughput measurement based on one-way delay provides more precise and faster results compared to those based on round-trip delay.
\end{proposition}

The works in \cite{zhou2021tsengine,zhou2021communication} also use model parameter traffic for probing link throughput, however, they employ the round-trip delay measurement with a stop-and-wait protocol. In contrast, our approach, based on one-way delay measurement, provides more precise and faster results, as proved in Proposition \ref{prop:measure-accuracy}. To further improve measurement accuracy, we implement two techniques: filtering tiny chunks and clock synchronization.

\textit{Filtering Tiny Chunks.} 
Deep learning models often contain tiny parameter tensors, such as biases and convolutional kernels, which introduce errors into link throughput measurements. For tiny chunks, the relative impact of message processing time is more significant. Additionally, these tiny chunks require less buffer space and thus can be processed and forwarded with higher priority. To mitigate these, a threshold, \texttt{PROBE\_CHUNK\_SIZE}, is introduced to balance measurement accuracy and timeliness. Specifically, only model chunks larger than this threshold are included in the calculation of link throughput as they provide reliable measurements, while tiny chunks are excluded. If set too high, most chunks will be excluded, leaving an insufficient number of chunks for timely measurement. Conversely, if set too low, many tiny chunks will be included, leading to errors in measurements. Thus, finding an optimal threshold is crucial to balance these considerations.

\textit{Clock Synchronization.} 
Proposition \ref{prop:measure-accuracy} is built on an assumption: the sender and receiver have synchronized clocks. If the clocks are not synchronized, the calculation of $t_{\mathrm{r}}^i-t_{\mathrm{s}}^i$ in Eq. \eqref{eq:throughput} will be impacted by clock drift. To combat this, as depicted in Figure \ref{fig:awareness-module}, a clock synchronization mechanism is implemented, featuring a server on the scheduler and a proxy on each node. The clock synchronization server runs a Network Time Protocol (NTP) daemon to align the clocks among root servers, after which the proxy on each parent node aligns the clocks of their child nodes, following the FAPTs determined by Algorithm \ref{alg:r-rooted-fapt-topo-construct}. Given that these FAPTs are designed with minimal transfer delay, they help minimize the clock drift across different nodes.
\section{Multipath Auxiliary Transmission}\label{sec:multipath-aux-transmission}
\begin{figure}[h]
\centering
\includegraphics[width=.48\textwidth]{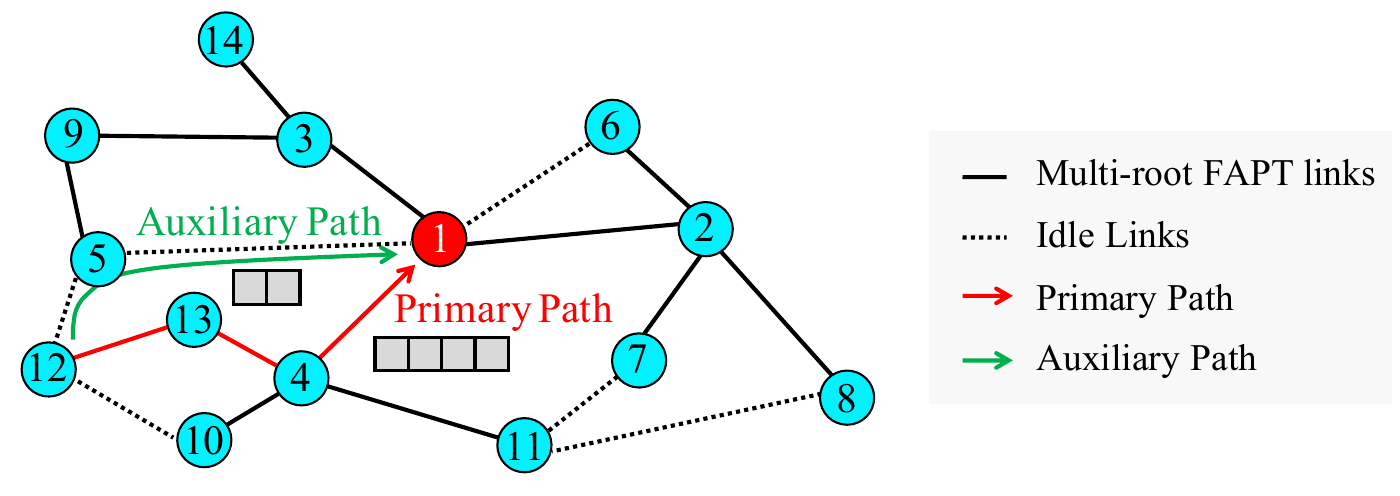}
\caption{An example of multipath auxiliary transmission.}
\label{fig:multi-path-auxiliary-routing}
\end{figure}

Although the network awareness module in Section \ref{sec:network-awareness} is effective, it encounters the significant challenge of the ``avalanche effect''. This effect occurs because the model parameter traffic used for probing is restricted within the existing synchronization topology, leaving external links unmeasured, as shown in Figure \ref{fig:multi-path-auxiliary-routing}. As a result, the network awareness module may only provide partial link information, risking suboptimal decisions in the synchronization topology. 

\subsection{Overall Design}
To overcome this, this section introduces a multipath auxiliary transmission mechanism that utilizes idle links to assist the primary paths in transmission. Here, we declare again that, the primary paths refer to the fastest aggregation paths within the determined multi-root FAPT synchronization topology, while the auxiliary paths are created from idle links. Typically, these auxiliary paths are slower, and thus, used to offload a small portion of model chunks from the primary paths. There can be multiple auxiliary paths between each pair of nodes. This design ensures that all links are involved in transmission and measurement, thus enhancing network awareness and further accelerating model transmission. Notably, nodes on the primary paths operate in ``aggregate-forward'' mode, while nodes on auxiliary paths operate in ``forward-only'' mode. This design prevents the slower auxiliary paths from causing unnecessary synchronization blockages and becoming bottlenecks that could hinder aggregation on the primary paths.

\begin{figure}[t]
\centering
\includegraphics[width=.49\textwidth]{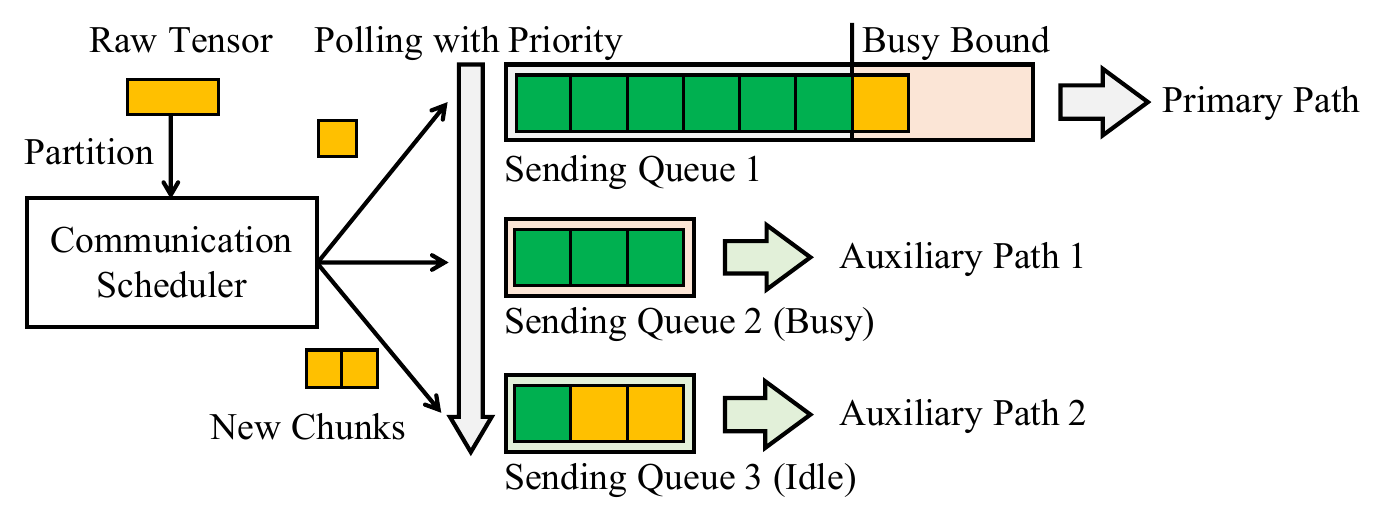}
\caption{The scheduling process for multipath auxiliary transmission. This case has one sending queue for the primary path and two for the auxiliary paths. Green squares are model chunks currently in transmission, whereas orange squares are recently arrived chunks.}
\label{fig:comm-scheduler}
\end{figure}

Figure \ref{fig:comm-scheduler} illustrates how the multipath auxiliary transmission is implemented. On the sender side, a communication scheduler is introduced to manage the transmission of model chunks. It partitions the raw parameter tensor into several similarly sized chunks and polls the primary and auxiliary paths in turn to assign the new chunks to their sending queues. Specifically, the communication scheduler defaults to assign the new chunks to the primary path, which acts as a ``data highway'', facilitating fast delivery of the majority of chunks. However, when the primary path becomes busy, that is, the occupied queue size exceeds \texttt{PRIMARY\_BUSY\_BOUND}, new chunks are assigned to an auxiliary path. These auxiliary paths are ranked by their transfer delay, with faster paths given higher priority. The communication scheduler iterates through these auxiliary paths in order of priority, assigning new chunks to an auxiliary path with occupied queue size below \texttt{AUXILIARY\_QUEUE\_LENGTH}. If all auxiliary paths are busy, it defaults back to using the primary path.

The above design uses two hyperparameters: \texttt{PRIMARY\_} \texttt{BUSY\_BOUND} and \texttt{AUXILIARY\_QUEUE\_LENGTH}. The first defines the threshold at which the primary path is considered busy: when the number of transmitting chunks exceeds this threshold. Upon reaching this threshold, the communication scheduler activates the auxiliary paths to assist with model chunk transmission. The latter specifies the capacity limit for transmitting model chunks on an auxiliary path. Once reaching this limit, that auxiliary path cannot accept more chunks until some of the ongoing transmissions are completed, requiring a shift to other available paths.

\subsection{Auxiliary Path Search Algorithm}
To find all non-overlapping paths between each pair of nodes, this section presents an auxiliary path search algorithm, outlined in Algorithm \ref{alg:aux-path-search}. \new{This algorithm is also executed by the scheduler, alongside Algorithm \ref{alg:r-rooted-fapt-topo-construct}.} It adopts an exclusionary strategy, where it iteratively runs Algorithm \ref{alg:find-all-fapt-trees} to identify the current shortest path set $H$ and then remove the used edges from the graph $G$, yielding a progressively pruned graph $G = G \setminus \{e | \forall e \in H\}$. This procedure is repeated on the pruned graph until no edges remain. The outcome is a set $ H_{\mathrm{aux}}^{i,j} $ containing several shortest paths from any node $v_i$ to another node $v_j$. In this set, the first path $ H_{\mathrm{aux}}^{i,j}(0) $ serves as the primary path, having minimal synchronization delay, while the subsequent paths $ H_{\mathrm{aux}}^{i,j}(i), \forall i \ge 1 $, represent the auxiliary paths, each offering faster delivery than the subsequent one.

\begin{algorithm}[t]
\caption{\textsc{Auxiliary Path Search}}
\label{alg:aux-path-search}
\begin{algorithmic}[1]
\small
\Require An undirected graph $G=(V,E)$; the data throughput $s(e)$ for each link $e$.
\Ensure Set of non-overlapping paths $H_{\mathrm{aux}}$ between any two nodes.
\State{Initialize the path set $H_{\mathrm{aux}}=\varnothing$;}
\While{there are unused edges $|E|>0$}
\State{Reset the transfer delay $w_{\mathrm{trans}}(e)$ for each link $e\in E$ using the latest link throughput $s(e)$: $w_{\mathrm{trans}}(e) = \frac{1}{s(e)}, \forall e \in E$;}
\State{$H\gets\textsc{Find-Fastest-Aggregation-Paths}(G,w_{\mathrm{trans}})$;}
\For{any two reachable nodes $v_i,v_j$}
\State{Retrieve the path $p_{i\rightarrow j}$ from set $H$;}
\State{Add $p_{i\rightarrow j}$ to set $H_{\mathrm{aux}}$: $H_{\mathrm{aux}}\gets H_{\mathrm{aux}}\cup\{p_{i\rightarrow j}\}$;}
\State{Remove used edges from $G$: $E\gets E\backslash\{e|\forall e\in p_{i\rightarrow j}\}$;}
\EndFor
\EndWhile
\State{\Return{$H_{\mathrm{aux}}$};}
\end{algorithmic}
\end{algorithm}

\textit{Complexity Analysis.} 
The time complexity of Algorithm \ref{alg:aux-path-search} mainly depends on the complexity of Algorithm \ref{alg:find-all-fapt-trees} and the number of times it is invoked. In the worst case, Algorithm \ref{alg:find-all-fapt-trees}, with a time complexity of $\mathcal{O}(|V^3|)$, will be invoked $\frac{|E|}{|V|-1}$ times. As a result, the overall complexity of Algorithm \ref{alg:aux-path-search} is $\mathcal{O}(|E||V|^2)$, where $|E|,|V|$ denote the total number of edges and nodes in graph $G$, respectively.
\section{Distributed Policy Consistency Protocol}
\label{sec:policy-consistency}
The parameter synchronization topology and auxiliary paths, collectively termed ``policy'', require periodic updates to enable \NAME~to adjust to network changes. When a new policy is formulated, it should be applied across all distributed nodes. However, the delivery of the new policy can be delayed, causing inconsistencies among nodes. This can lead to operational conflicts: if some nodes switch to the new policy while others remain on the old one, it might result in deadlocks and packet loss. 

\begin{figure}[h]
\centering
\includegraphics[width=0.48\textwidth]{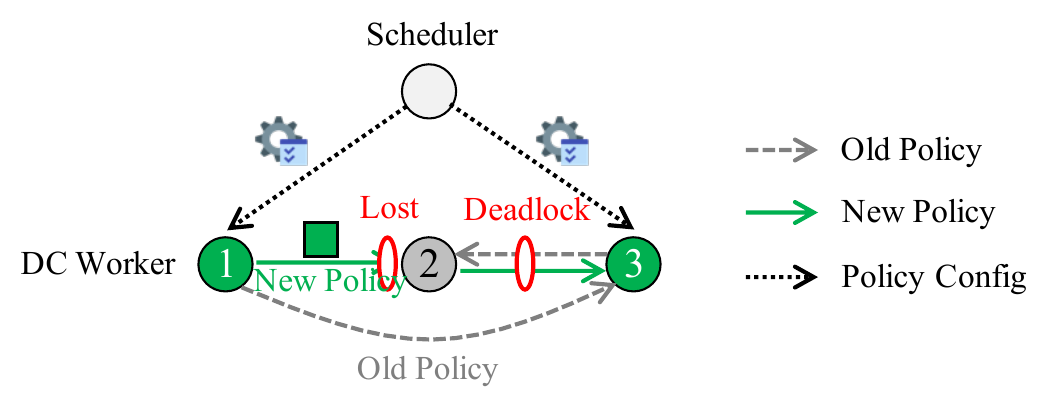}
\caption{An illustration of policy inconsistency leading to packet loss and deadlock.}
\label{fig:inconsistency}
\end{figure}

As depicted in Figure \ref{fig:inconsistency}, nodes 1 and 3 have switched to the new policy while node 2 lags behind, causing coexistence of both old and new policies. Under such a condition, node 2, adhering to the old policy, expects data from node 3, but node 3, following the new policy, awaits data from node 2, leading to a deadlock. Furthermore, node 1, under the new policy, sends data to node 2, but node 2, under the old policy, does not recognize this incoming data, leading to packet loss. 

To address these risks, this section designs a robust policy consistency protocol to ensure seamless transitions between old and new policies. Given the different traffic behaviors on primary and auxiliary paths, where primary path traffic operates in ``aggregate-forward'' mode and auxiliary path traffic in ``forward-only'' mode, separate consistency protocols are designed for each.

\subsection{Synchronization Topology Consistency Protocol}
\begin{figure}[t]
\centering
\includegraphics[width=0.49\textwidth]{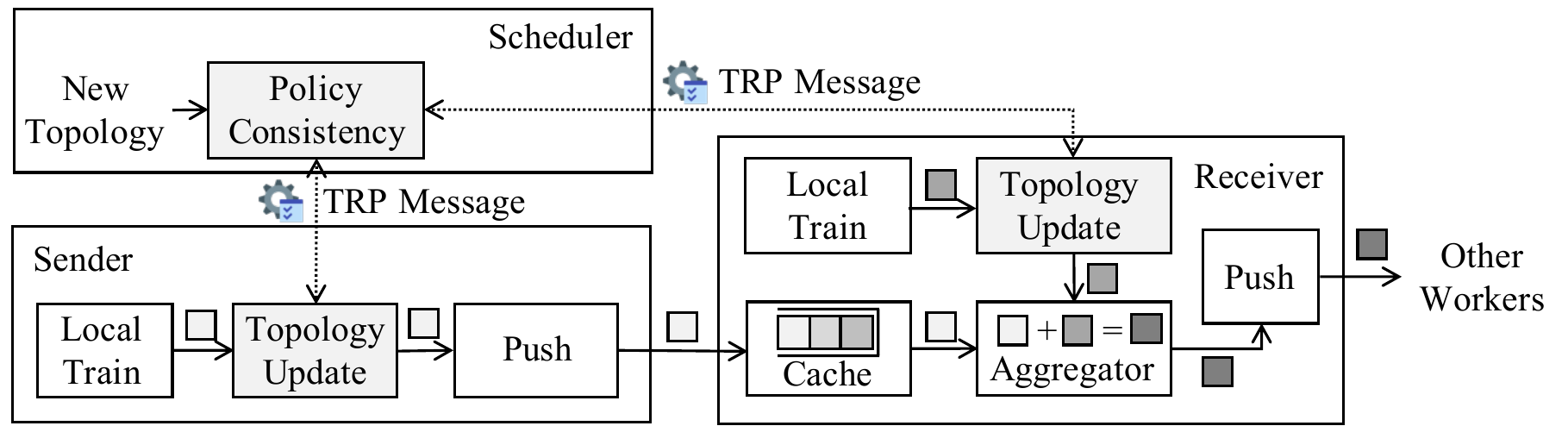}
\caption{The synchronization topology consistency protocol workflow.}
\label{fig:trp-protocol}
\end{figure}

The consistency protocol for parameter synchronization topology is depicted in Figure \ref{fig:trp-protocol}. In this protocol, worker nodes initiate topology update requests using a Topology Request Protocol (TRP) message sent to the scheduler before each \texttt{PUSH} operation. Then, they enter a blocking state awaiting a response, which could be either a topology update or a confirmation that no update is required. The feasibility of this protocol is demonstrated through the following cases.

\begin{case}
This protocol guarantees that all worker nodes have their local topology updated before initiating communication. In other words, when a worker node starts to transmit model data, it must have the most recent synchronization topology, ensuring that the model data is sent to the correct receiver.
\end{case}

\begin{case}
When a worker node receives model data but its local topology has not been updated, it will cache the received data and then process it once the local topology has been updated, thus ensuring that any model data arriving early is preserved and not discarded.
\end{case}

\subsection{Auxiliary Path Consistency Protocol}
\begin{figure}[h]
\centering
\includegraphics[width=0.45\textwidth]{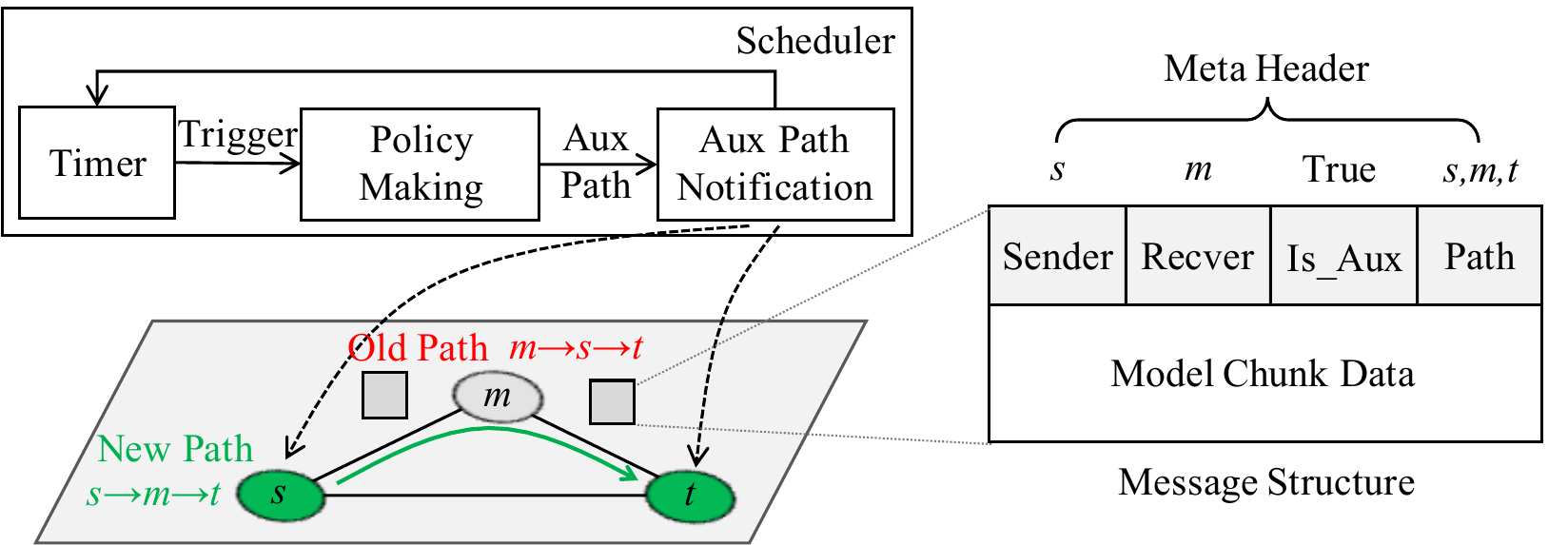}
\caption{The consistency protocol and its message for auxiliary paths.}
\label{fig:route-consistency}
\end{figure}

The consistency protocol for auxiliary paths is illustrated in Figure \ref{fig:route-consistency}. To ensure the timeliness of auxiliary paths, the policy formulation module is triggered every \texttt{UPDATE\_TIME} seconds to refresh the auxiliary paths, and then the new paths are distributed to all worker nodes. However, notification delays can lead to the temporary coexistence of old and new auxiliary paths. As shown in Figure \ref{fig:route-consistency}, while nodes $s$ and $t$ have switched to the new auxiliary path $s\rightarrow m\rightarrow t$, node $m$ may still use the old path $m\rightarrow s\rightarrow t$, causing a network loop between $s$ and $m$.

To tackle this issue, a new message structure is introduced, which records the node sequence of the auxiliary path in the meta header. For example, when node $m$ receives model data, it first checks the \texttt{IS\_AUX} metadata. If this is set to \texttt{True}, the packet is confirmed to be on an auxiliary path. Then, node $m$ reads the \texttt{PATH} metadata, for example, $\{s, m, t\}$, to pinpoint the next destination node $t$. This design ensures that the forwarding along an auxiliary path is solely determined by the source node, regardless of the intermediate nodes’ auxiliary paths. In this way, even if the source node is using an outdated auxiliary path, data transmission remains successful. \new{Importantly, since \NAME~operates at the application layer, all messages discussed in this paper are payloads encapsulated within TCP/IP packets.}
\section{System Overview}
\begin{figure}[h]
\centering
\includegraphics[width=0.4\textwidth]{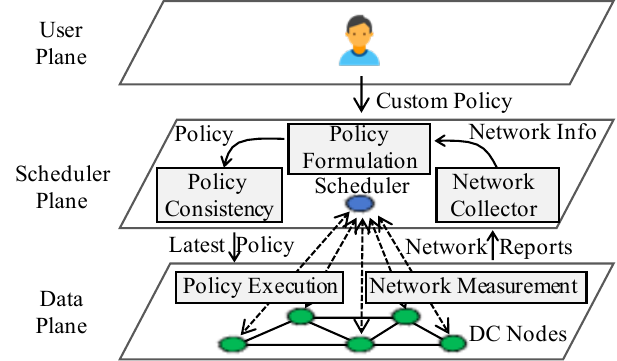}
\caption{The overview of our system architecture.}
\label{fig:system-overview}
\end{figure}

The comprehensive architecture of \NAME, depicted in Figure \ref{fig:system-overview}, is structured into three planes: the data plane, the scheduler plane, and the user plane.

\subsection{Data Plane}
The data plane is tasked with training ML models and synchronizing the learned parameters. It is made up of data center nodes that are interconnected through an overlay network. Within our system, nodes are engaged not only in model training and executing \texttt{PUSH} and \texttt{PULL} operations but also in receiving and aggregating model parameters from other nodes. Furthermore, these nodes have the capability to dynamically schedule synchronization traffic and monitor network conditions. To achieve these functions, the data plane is equipped with two modules: a policy execution module and a network measurement module.

\textit{Policy Execution Module.} This module retrieves the latest policy (i.e., the synchronization topology and auxiliary paths) issued by the scheduler and executes it on the worker node. It guides the worker node in identifying the correct sources and destinations for model data exchange.

\textit{Network Measurement Module.} This module measures the link throughput between the node and its neighboring nodes, then reports the results to the scheduler. This feedback mechanism realizes the network awareness functionality detailed in Section \ref{sec:network-awareness}, aiding the scheduler in making accurate decisions.

\subsection{Scheduler Plane}
This plane is the core of \NAME, containing only a scheduler that can be colocated with any worker node to orchestrate the parameter synchronization process in the data plane. Specifically, this plane features a network collector module to gather network reports from the data plane. Using this information, the scheduler can reformulate the synchronization topology and auxiliary paths, and issue the new policy to worker nodes for execution. To achieve these functions, the scheduler is implemented with three modules: network collector, policy formulation, and policy consistency.

\textit{Network Collector Module.} This module continuously collects network reports from the data plane, tracking any network changes and initiating the policy formulation module when significant changes are detected. In our implementation, the threshold for recognizing a significant change is set to 0. This means that the synchronization topology and auxiliary paths are refreshed every \texttt{UPDATE\_TIME} seconds, regardless of whether a significant change has occurred.

\new{\textit{Policy Formulation Module.} Utilizing the most current network reports from the network collector, this module employs Algorithm \ref{alg:r-rooted-fapt-topo-construct} to formulate the synchronization topology and Algorithm \ref{alg:aux-path-search} to search for the auxiliary paths. Given the dynamic nature of network conditions, this module is periodically invoked to refresh the policy.}

\textit{Policy Consistency Module.} The new policy needs to be synchronized to worker nodes to update their local policies. This module manages this task, and guarantees a seamless transition between the old and new policies via two consistency protocols elaborated in Section \ref{sec:policy-consistency}. 

\subsection{User Plane}
\NAME~defaults to use Algorithm \ref{alg:r-rooted-fapt-topo-construct} for automatic synchronization topology decision. However, users may need to customize the topology to better align with their needs. This plane allows for customization of the synchronization topology and provides options for users to customize \NAME~behaviors, as detailed in Table \ref{table:customizable-options}.

\begin{table}[h]
\renewcommand\arraystretch{1.2}
\centering
\caption{Customizable options in \NAME~for users.}
\begin{tabular}{c|l}
\hline
\textbf{Option} & \textbf{Description} \\ 
\hline\hline
ENABLE\_AWARENESS & Whether to enable network awareness. \\
ENABLE\_AUX\_PATH & Whether to use auxiliary paths. \\
UPDATE\_RATE & Threshold to distinguish significant changes. \\
\hline
\end{tabular}
\label{table:customizable-options}
\end{table}
\section{Experiments and Analysis}
We built \NAME~on top of MXNET \cite{chen2015mxnet}, refactoring the KVSTORE and PS-LITE modules. In MXNET, nodes are classified as roles, including workers, servers, and a scheduler. The worker nodes only train the local model, and the servers only handle parameter synchronization. \NAME, however, introduces a general node that combines the functionalities of both workers and servers. These general nodes are capable of training models and managing parameter synchronization operations, including the reception, pushing, forwarding, and aggregation of model data, from or to any other general nodes. They are also equipped with essential features for network measurement and traffic scheduling. Moreover, the scheduler in \NAME~has been enhanced to support network awareness, policy formulation, and policy consistency. To validate our prototype, this section conducted comprehensive experiments, including comparative analysis, ablation studies, hyperparameter optimization, and scalability tests.

\subsection{Testbed Setup}
\begin{figure}[t]
\centering
\includegraphics[width=0.48\textwidth]{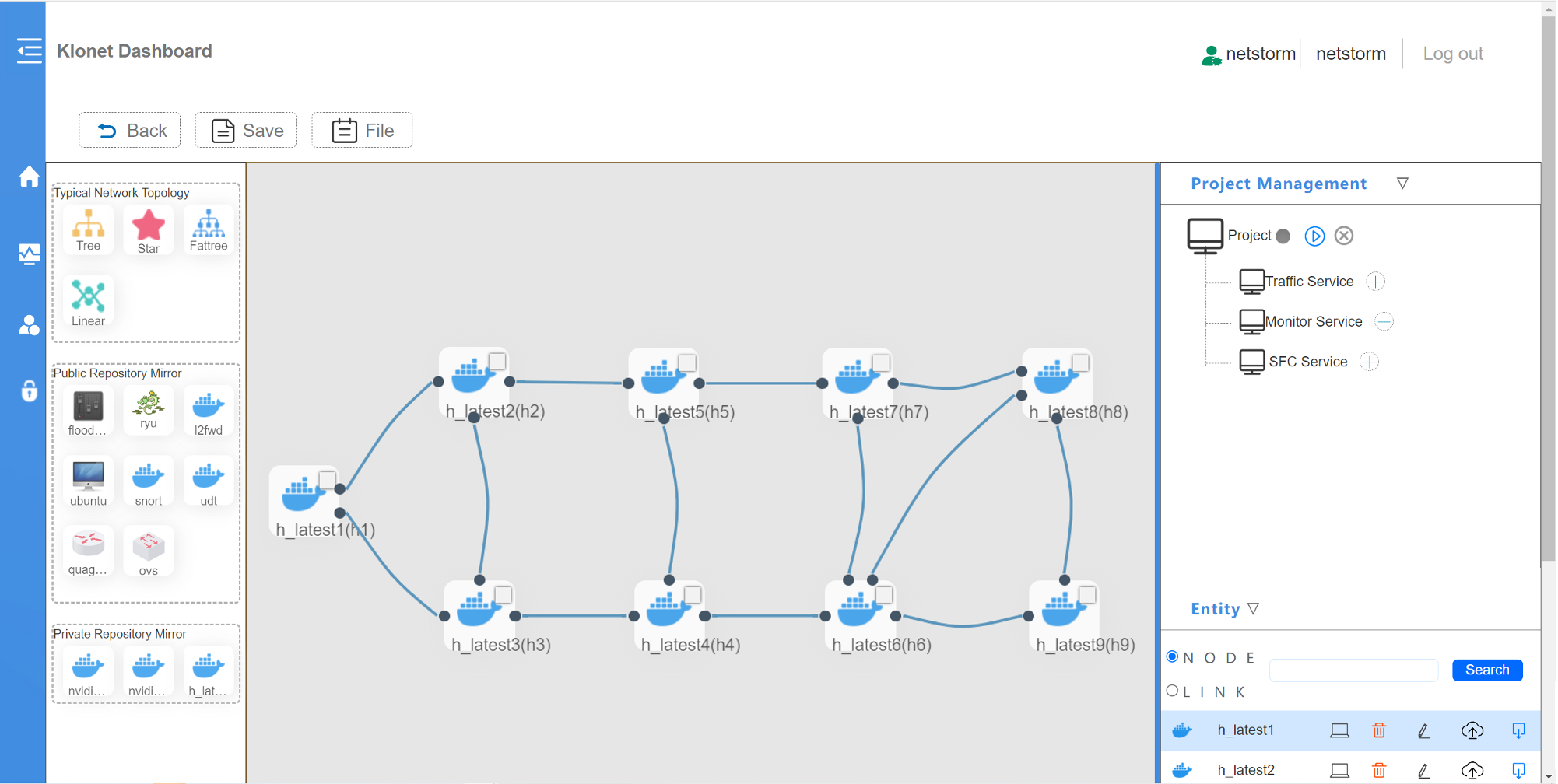}
\caption{Overlay network topology emulated by Klonet.}
\label{fig:klonet}
\end{figure}

Our hardware platform includes 10 Intel Xeon CPUs and 8 2080TI/3090 GPUs, totaling 240 logical cores, 880 GB of memory, and 140 GB of VRAM. \new{This hardware supplies the requisite physical resources for Klonet \cite{ma2024klonet}, a large-scale network emulation platform, to create a realistic WAN environment. Klonet supports network emulations with layer two and above, allowing network properties such as overlay network topology, bandwidth, latency, and packet loss on virtual links to be configured. This capability makes Klonet ideal for emulating a wide range of network scenarios, particularly WANs. As a result, the WAN network emulated by Klonet shares the same properties as an actual WAN network.} Figure \ref{fig:klonet} illustrates an overlay network topology emulated by Klonet, which is a simplified version of a widely used WAN backbone, Internet2. \new{This topology consists of 9 data centers located in different regions, with a latency of 30ms and a packet loss rate of 0.02\%. By adjusting the available bandwidth on virtual links, we set the data rates on these links ranging from 20 to 155 Mbps.} Furthermore, to reflect resource dynamics in WANs, the data rates change every 3 minutes.

\NAME~optimizes parameter synchronization at the network level, ensuring that the training convergence property remains unaffected. This allows for flexible selection of the training model, dataset, and optimizer for our experiments. Thus, we choose the Fashion-MNIST dataset, and distribute the training data evenly across data centers. Adam optimizer is used to train the AlexNet model, with a learning rate of 1e-4 and a batch size of 32. Then, we focus our sights on evaluating the factors that affect synchronization efficiency. The default settings for these factors are detailed in Table \ref{table:summary-of-hyper-params}.

\begin{table}[h]
\renewcommand\arraystretch{1.2}
\centering
\caption{Summary of default hyperparameter setup.}
\begin{tabular}{c|c|c}
\hline
\textbf{Variable} & \textbf{Symbol} & \textbf{Value} \\ 
\hline\hline
NUM\_NODES & $|V|$ & 9 \\ \hline
NUM\_ROOT\_SERVERS & $N$ & 9 \\ \hline
CHUNK\_SIZE & - & 1 million \\ \hline
PRIMARY\_BUSY\_BOUND & - & 2 \\ \hline
AUXILIARY\_QUEUE\_LENGTH & - & 1 \\ \hline
PROBE\_CHUNK\_SIZE & - & 2 millions \\ \hline
PROBE\_CHUNK\_NUM & $I$ & 4 \\ \hline
UPDATE\_TIME & - & 5 seconds \\ \hline
\end{tabular}
\label{table:summary-of-hyper-params}
\end{table}

\subsection{Comparative Analysis}
To evaluate the effectiveness of \NAME, we compare it with three alternatives: MXNET \cite{chen2015mxnet}, MLNET \cite{mai2015optimizing}, and TSEngine \cite{zhou2021tsengine}.
\begin{itemize}
    \item \textbf{MXNET} is a distributed training system widely adopted in GeoML domains, built upon the standard PS architecture. In such a system, data center nodes directly communicate with a central center, forming a starlike overlay topology known as Hub-and-Spokes. 
    \item \textbf{MLNET} employs a balanced $k$-way tree (BKT) structure for its synchronization topology, aiming to speed up model synchronization. In this study, we evaluate this topology structure in WAN environments to assess its training efficiency.
    \item \textbf{TSEngine}, also utilizing a tree-based synchronization topology, aims to construct a minimum spanning tree (MST). In this system, nodes prefer selecting links with the highest data throughput for data transmission, thus avoiding bottleneck links.
\end{itemize}

\begin{figure}[h]
\centering
\includegraphics[width=0.5\textwidth]{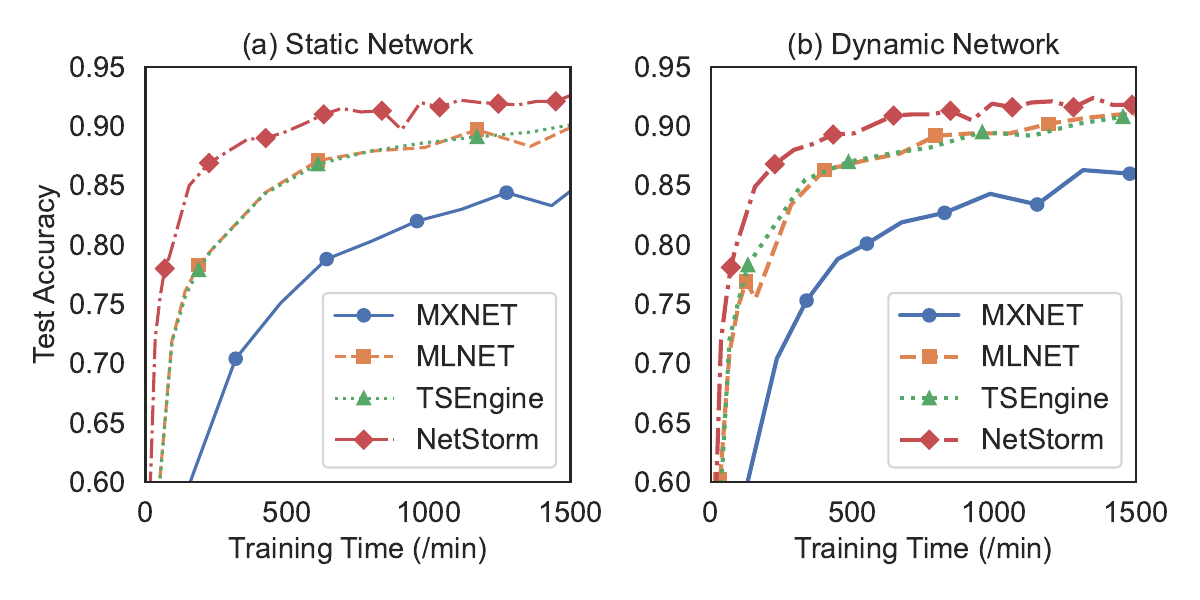}
\caption{The training efficiency for \NAME~and baseline systems.}
\label{fig:system-comparison}
\end{figure}

The performance comparison of \NAME~with MXNET, MLNET, and TSEngine is shown in Figure \ref{fig:system-comparison}. In both static and dynamic networks, \NAME~consistently exhibits superior training speeds over the other systems. Leveraging its multi-root FAPT topology and multipath auxiliary transmission mechanism, under stable network conditions, \NAME~achieves training speeds up to 9.2 times faster than MXNET. Even in environments with frequent network changes, \NAME~can dynamically adjust its synchronization topology and auxiliary paths, thus maintaining a training acceleration factor of 6.5 times. These results highlight \NAME's substantial benefits in speeding up GeoML training and showcase its strong adaptability to network heterogeneity and dynamics. Notably, the training times recorded in this experiment include model evaluation time, which will be excluded in subsequent experiments.

\subsection{Ablation Analysis}
Before proceeding with the ablation analysis, we define the \textit{normalized data throughput}, which refers to the ratio of the total number of training samples processed per second by this evaluated system compared to that of MXNET. This metric indicates the changes in parameter synchronization delay, where \textit{a higher normalized data throughput typically leads to a lower parameter synchronization delay}. With this definition, ablation studies are carried out to evaluate the individual contributions of each component, including the FAPT topology, its multi-root variant, the network awareness module, and the multipath auxiliary transmission mechanism. Each set of experiments is repeated 5 times to ensure reliability.

\begin{figure}[h]
\centering
\includegraphics[width=0.48\textwidth]{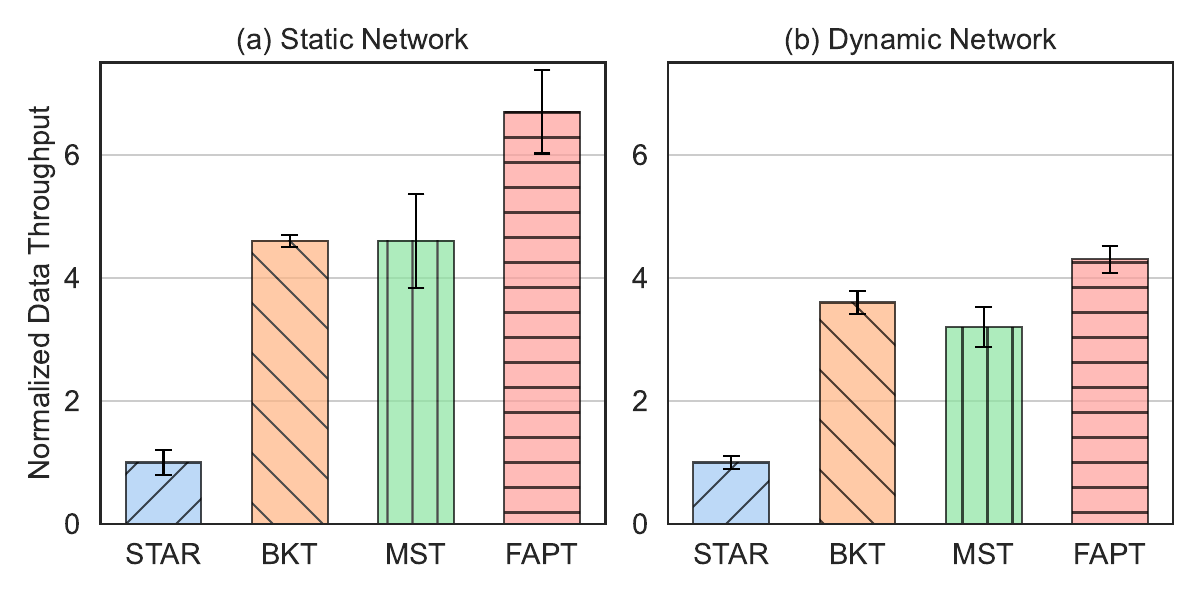}
\caption{The normalized throughput for different overlay topologies.}
\label{fig:topo-comparison}
\end{figure}

\textbf{(1) Use FAPT Topology to Maximize Throughput.}
We start by comparing the synchronization topologies used by MXNET, MLNET, TSEngine, and \NAME, specifically STAR, BKT, MST, and FAPT, respectively. STAR and BKT feature fixed regular structures, whereas MST and FAPT feature adaptable irregular tree structures. To ensure a fair comparison, FAPT was configured with only one root. The comparison of their normalized data throughput is shown in Figure \ref{fig:topo-comparison}. Tree-based topologies demonstrate superior performance over the star structure in both static and dynamic networks, with FAPT standing out as the most effective. This result aligns well with Figure \ref{fig:tree-topos}f and highlights the importance of choosing an appropriate synchronization topology. The right choice can bring significant speedup, up to 6.7 times faster. Moreover, the topology design should consider the heterogeneity of network resources. FAPT achieves this, offering a 210\% speedup over regular balanced trees. Lastly, the ability to adaptively adjust the topology in line with network dynamics is also crucial. An adaptive FAPT yields a 70\% speedup compared to a static balanced tree in a dynamic network.

\begin{figure}[t]
\centering
\includegraphics[width=0.3\textwidth]{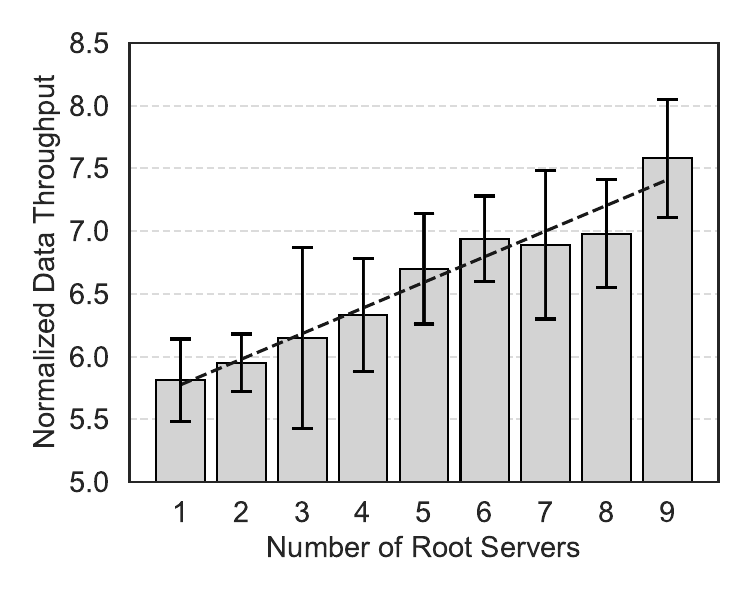}
\caption{The normalized throughput with increasing root servers.}
\label{fig:num-root-load-balance}
\end{figure}

\textbf{(2) Accelerate FAPT Topology with More Root Servers.}
Now, we increase the number of root servers to assess their effect in a dynamic network, with the multipath auxiliary transmission activated. As shown in Figure \ref{fig:num-root-load-balance}, the normalized data throughput increased linearly with more roots, benefiting from better load balancing and increased transmission parallelism, which can be intuitively inferred from Figure \ref{fig:multi-server-load-balance}. Specifically, the peak throughput was attained when the number of roots matching data center nodes, that is, each data center node functions as a root, leading to a 180\% speedup compared to having only one root.

\begin{figure}[t]
\centering
\includegraphics[width=0.3\textwidth]{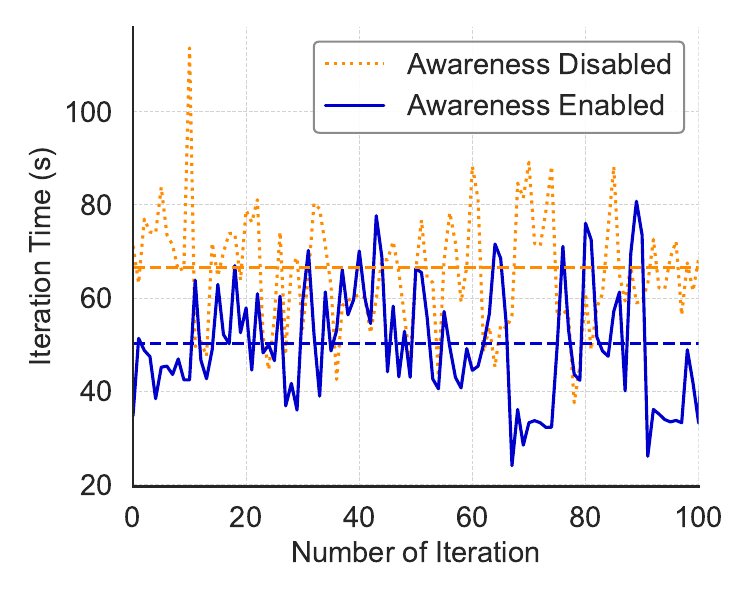}
\caption{Iteration time with network awareness enabled and disabled.}
\label{fig:enable-awareness}
\end{figure}

\new{\textbf{(3) Enable Network Awareness to Adapt to Dynamic Networks.}
Network awareness allows our system to dynamically adjust its synchronization topology and auxiliary paths to align with the most recent network conditions, ensuring the transmission policy is always up-to-date. In this experiment, the network fluctuates over time, and the system performance was assessed by comparing iteration times with and without network awareness. As shown in Figure \ref{fig:enable-awareness}, without this feature, the average iteration time stands at 66 seconds, however, with network awareness activated, it drops to 50 seconds. This reduction translates to a 24\% improvement in training speed, underscoring \NAME’s effectiveness and robustness, with network awareness, to adapt to WAN dynamics.}

\textbf{(4) Use Auxiliary Paths for Multipath Transmission.}
In this experiment, we activate network awareness and configure the multi-root FAPT topology with 9 root servers to assess the performance gains brought by the multipath auxiliary transmission mechanism. \new{For ease of explanation, we define an \textit{normalized throughput gain} as follows:
\begin{equation}
\texttt{Gain}=\frac{\tau_{\texttt{aux-path-enabled}}}{\tau_{\texttt{aux-path-disabled}}}-1
\end{equation}
Here, $\tau_{\texttt{aux-path-enabled}}$ represents the training data throughput per second with multipath auxiliary transmission activated, while $\tau_{\texttt{aux-path-disabled}}$ is not. Generally, a higher value of \texttt{Gain} indicates a shorter training time.}

\begin{figure}[t]
\centering
\includegraphics[width=0.317\textwidth]{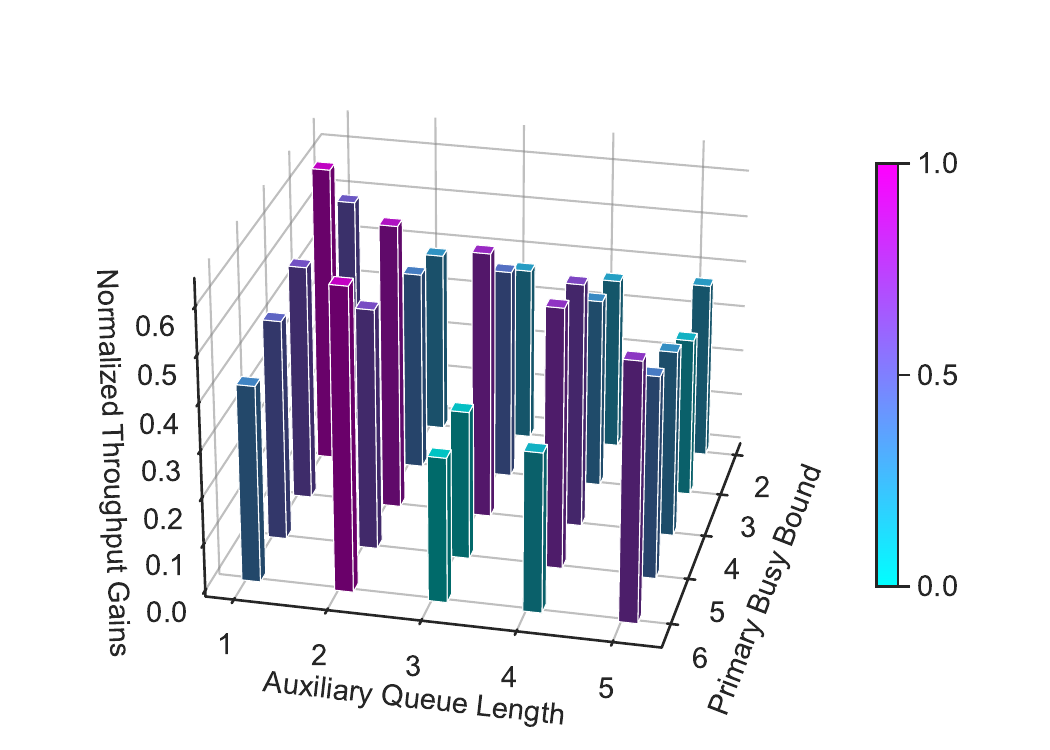}
\caption{Throughput gains with multipath auxiliary routing activated.}
\label{fig:auxiliary-routing}
\end{figure}

\new{The performance of multipath auxiliary transmission is influenced by two factors: \texttt{PRIMARY\_BUSY\_BOUND} and \texttt{AUXILIARY\_QUEUE\_LENGTH}. Therefore, we carried out a grid search over various hyperparameter combinations, as shown in Fig. \ref{fig:auxiliary-routing}. The results indicate that across all hyperparameter settings, the gain spans from 30\% to 65\%, with the optimal configuration being \texttt{PRIMARY\_BUSY\_BOUND=2} and \texttt{AUXILIARY\_QUEUE\_LENGTH=5}. This suggests that this feature can reduce the training time by 40\%.}

In the optimal setup, the primary path is ``busy'' when it has 2 model chunks in transit, triggering the use of auxiliary paths. Then, if all the auxiliary paths have 5 model chunks in transit, further chunks will be directed back to the primary path. Therefore, although \texttt{AUXILIARY\_QUEUE\_LENGTH} can be set higher than \texttt{PRIMARY\_BUSY\_BOUND}, the primary path still handles most of the model chunk transmission. By engaging auxiliary paths sooner, \NAME~can make full use of idle links for dispersing network traffic and mitigating congestion in hotspots, and more importantly, enhancing network awareness by incorporating these idle links into measurements.

\begin{figure}[t]
\centering
\includegraphics[width=0.295\textwidth]{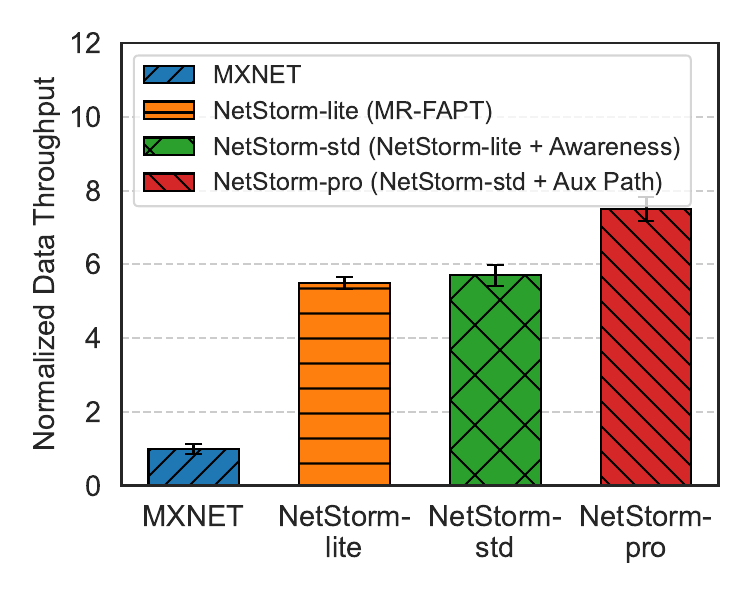}
\caption{An overview of ablation studies for each technique.}
\label{fig:gain-analysis}
\end{figure}

\textbf{(5) Integrate Them into \NAME.} 
Next, we integrate the aforementioned technologies and overview their combined effect. Taking MXNET as the baseline, we activate the multi-root FAPT topology (MR-FAPT), network awareness, and multipath auxiliary transmission in order, naming these stages \texttt{NetStorm-lite}, \texttt{NetStorm-std}, and \texttt{NetStorm-pro}, respectively. As Figure \ref{fig:gain-analysis} shows, with MR-FAPT alone, \texttt{NetStorm-lite} attains a training speed 5.5 times faster than MXNET, highlighting the efficacy of our synchronization topology. Then, by incorporating network awareness, \texttt{NetStorm-std} allows dynamic adaptation of the synchronization topology to varying network conditions, yielding a 20\% increase in throughput. Notably, this gain is somewhat limited because auxiliary paths are not yet used nor measured, resulting in incomplete network knowledge. Finally, we activate multipath auxiliary transmission in \texttt{NetStorm-pro} to fix this issue, which achieves enhanced network awareness and increased transmission parallelism, culminating in a 7.5-fold speedup over the baseline.

In summary, each technology has an incremental contribution and provides essential features like network awareness, multipath transmission, multi-root load balancing, and policy consistency. They combine to substantially reduce parameter synchronization delays, making \NAME~a competitive solution for GeoML applications.

\subsection{Scalability Analysis}
\begin{figure}[t]
\centering
\subfloat[Increasing Model Size]{\includegraphics[height=3.95cm]{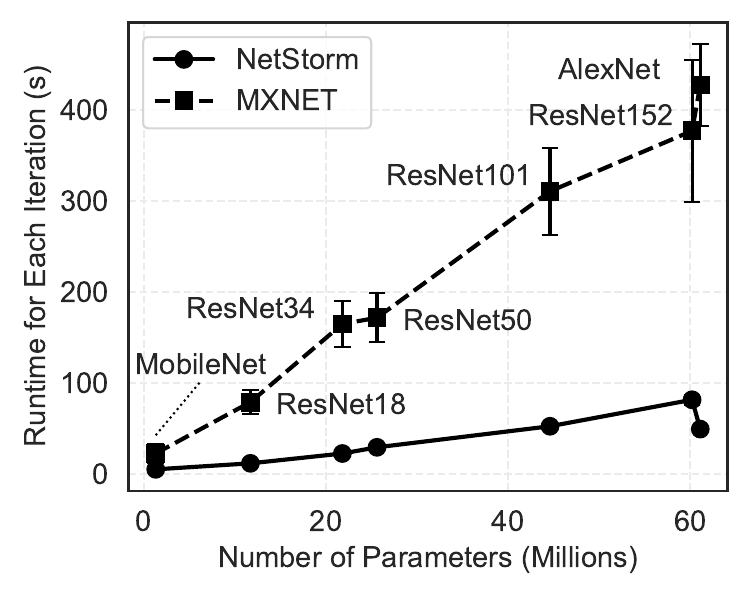}\label{fig:model-scalability}}
\subfloat[Increasing Cluster Size]{\includegraphics[height=4cm]{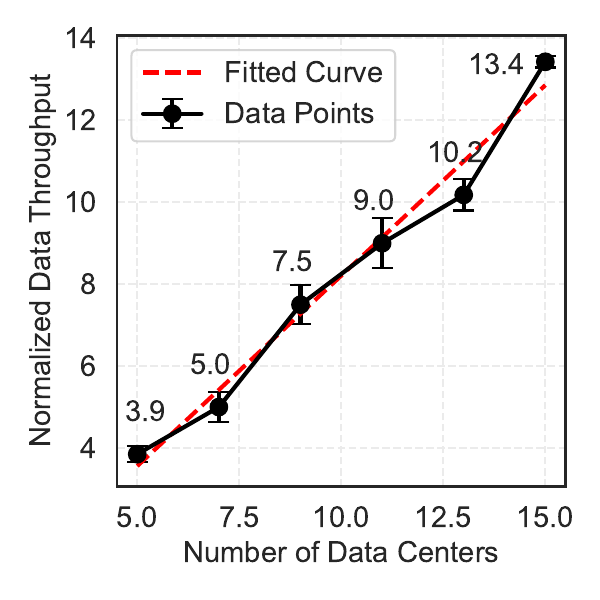}\label{fig:cluster-scalability}}
\caption{Scalability with increasing model size and cluster size.}
\end{figure}

\textbf{Scaling Up with Model Size.}
This experiment further explores the scalability in handling ML models with varying numbers of parameters. Specifically, we evaluate models like MobileNet, AlexNet, and various ResNet variants, with their average iteration runtimes presented in Figure \ref{fig:model-scalability}. As expected, both MXNET and \NAME~experience increased synchronization delays when processing larger-scale models, which is intuitive since more and larger parameter tensors require longer transmission times. However, \NAME~shows a notably more gradual rise in delay compared to MXNET, suggesting superior scalability in relation to model size.

An easter egg was observed with AlexNet: despite having a similar number of model parameters to ResNet-152, \NAME~achieves greater efficiency in training AlexNet. This efficiency arises from fully connected (FC) layers. With most of the model parameters residing on FC layers, the parameter tensors can be sliced more evenly, leading to a more balanced traffic distribution. This observation uncovers \NAME’s potential in handling large models, particularly Transformer models, which mainly consist of FC layers.

\new{\textbf{Larger Cluster Size Leads to Greater Speedup.}
We also evaluate the scalability on clusters of varying sizes. In this experiment, we increase the number of data centers from 5 to 15 and set the number of root servers to align with it. The results, as shown in Figure \ref{fig:cluster-scalability}, showcase a linear increase in normalized data throughput as the cluster size grows, achieving a scaling efficiency of 0.93. This demonstrates that \NAME~consistently maintains a high training speed, even at larger cluster sizes.}

\begin{figure*}[t]
\centering
\includegraphics[width=\textwidth]{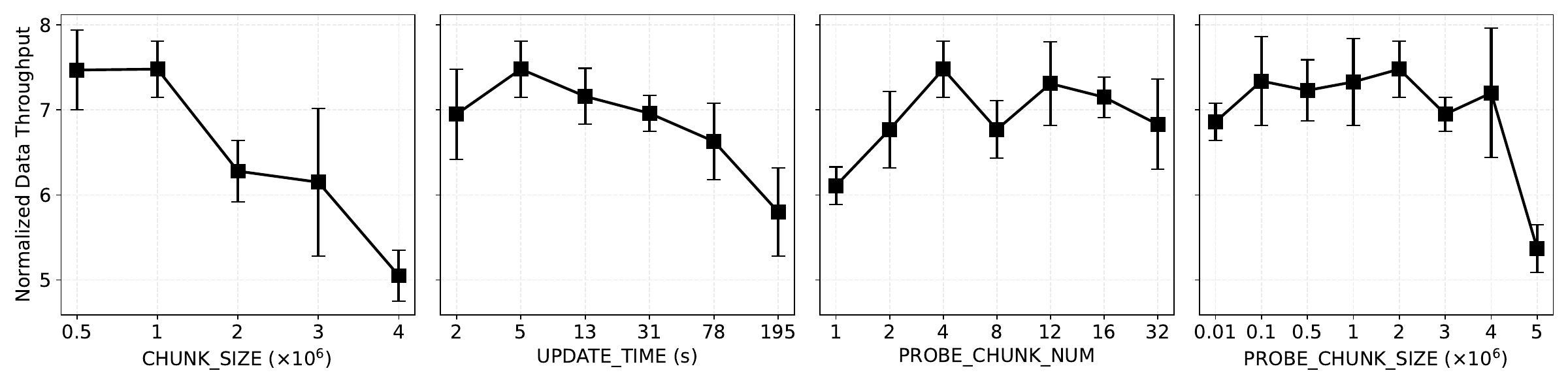}
\caption{Optimization of sensitivity hyperparameters in \NAME.}
\label{fig:sensitivity-search}
\end{figure*}

\subsection{Sensitivity Analysis}\label{subsec:sensitivity-analysis}
In Figure \ref{fig:sensitivity-search}, we test the effects of four hyperparameters: \texttt{CHUNK\_SIZE}, \texttt{UPDATE\_TIME}, \texttt{PROBE\_CHUNK\_SIZE}, and \texttt{PROBE\_CHUNK\_NUM}. The parameter most sensitive to value changes is \texttt{CHUNK\_SIZE}, which sets the maximum size for each model chunk. It was observed that increasing this value results in a significant decrease in normalized data throughput. The reasons are threefold: (a) larger chunks are more likely to cause congestion; (b) with model chunks growing in size, their number decreases, leaving some auxiliary paths without chunks assigned and some idle links remain unutilized and unmeasured; and lastly, (c) imbalanced loading, because tensors are sliced coarsely, causing significant deviations in parameter volume across root servers, sometimes leaving roots without any parameters to manage. Conversely, smaller model chunks enable more efficient bandwidth utilization and load balancing, and also facilitate more comprehensive network awareness. Therefore, a value between 0.5$\sim$1 million parameters is recommended, as it achieved the optimal speedup in this study.

The second sensitive parameter is \texttt{UPDATE\_TIME}, which sets the time interval for the scheduler to reformulate the synchronization topology and auxiliary paths. Generally, a larger value decreases the update frequency, making the system less responsive to network changes. Conversely, if set too low, the scheduler will frequently update and synchronize policies, incurring excessive overhead. This study identified an optimal setting of 5 seconds, at which \NAME~achieves a balance between sensitivity and update costs.

The number of probe chunks, \texttt{PROBE\_CHUNK\_NUM}, and the minimum size of these chunks, \texttt{PROBE\_CHUNK\_SIZE}, are two parameters associated with link throughput measurement. While they are less sensitive, proper calibration is still needed to prevent a sudden drop in training efficiency. Specifically, an overly high \texttt{PROBE\_CHUNK\_NUM} can cause increased latency in measurements, while too low a value might lead to significant measurement errors. A range of 4 to 32 is recommended to balance them. Notably, if \texttt{CHUNK\_SIZE} is set to be large, \texttt{PROBE\_CHUNK\_NUM} should be reduced accordingly.

For \texttt{PROBE\_CHUNK\_SIZE}, if set too low, more tiny chunks would be included for link throughput measurement, causing more measurement errors. Conversely, too large a value would exclude most model chunks, resulting in high measurement delays and a sudden drop in training efficiency. This ``collapse boundary'' is determined by both \texttt{CHUNK\_SIZE} and \texttt{PROBE\_CHUNK\_SIZE}. Generally, larger parameter tensors with a larger \texttt{CHUNK\_SIZE} correspond to a higher boundary value. Nonetheless, an excessively large \texttt{CHUNK\_SIZE} can also cause a drastic decline. Therefore, it is advisable to set both \texttt{CHUNK\_SIZE} and \texttt{PROBE\_CHUNK\_SIZE} to smaller values. For \texttt{PROBE\_CHUNK\_SIZE}, a range of 0.01 to 4 million parameters is recommended, as it has been found to yield greater efficiency.
\section{Conclusion}
This paper presents \NAME, a communication scheduler designed to speed up parameter synchronization across geo-distributed data centers, which often face challenges due to limited, heterogeneous, and dynamic network resources. We highlight the need for an effective metric for parameter synchronization topology, and design a multi-root FAPT topology, resulting in a 6.7-fold speedup. To acquire network knowledge, a passive network awareness module utilizing native model parameter probes has been implemented, leading to a 20\% increase in training speed within dynamic networks. To address the avalanche issue where some links remain unmeasured, we introduce a multipath auxiliary transmission mechanism. This mechanism utilizes idle links to establish auxiliary paths that assist the primary path in transmission, yielding a 65\% improvement in training speed. Lastly, we design distributed policy consistency protocols to ensure seamless policy updates. \NAME~was implemented and tested on Klonet. The results demonstrate that \NAME~significantly outperforms distributed training systems like MXNET, MLNET, and TSEngine, achieving an overall training speed of 6.5$\sim$9.2 times that of MXNET. Further analysis confirms \NAME’s scalability on varying cluster and model sizes.

\bibliographystyle{IEEEtran}
\begin{small}
\bibliography{bibliography}
\end{small}

\newpage
\appendices
\renewcommand{\theequation}{A.\arabic{equation}}
\setcounter{equation}{0}

\section{Proof of Theorem \ref{theorem:tree-completion-time}}
\begin{proof}
In a tree structure, consider any two leaf nodes \(v_i\) and \(v_j\), with paths to the root node \(v_r\) denoted as \(p_i\) and \(p_j\) respectively. Following Definition \ref{def:path-delay}, and after simplifying the processing delays, the overall synchronization delays for these paths, \(w(p_i)\) and \(w(p_j)\), can be represented as follows:
\begin{align}
w(p_i)&=\sum_{e\in p_i}{w_{\mathrm{trans}}^{i\rightarrow r}}(e)+\sum_{v_i\in p_i\backslash \{v_i\}}{w_{\mathrm{block}}^{i\rightarrow r}}(v_i),\\
w(p_j)&=\sum_{e\in p_j}{w_{\mathrm{trans}}^{j\rightarrow r}}(e)+\sum_{v_j\in p_j\backslash \{v_j\}}{w_{\mathrm{block}}^{j\rightarrow r}}(v_j).
\end{align}
Here, \(\sum_{e\in p_i}{w_{\mathrm{trans}}^{i\rightarrow r}}(e)\) denotes the total parameter transfer delay along the path from leaf node \(v_i\) to the root node \(v_r\), while \(\sum_{v_i\in p_i\backslash \{v_i\}}{w_{\mathrm{block}}^{i\rightarrow r}}(v_i)\) denotes the cumulative blockage delay for all non-leaf nodes on the path from leaf node \(v_i\) to the root node \(v_r\). \(\sum_{e\in p_j}{w_{\mathrm{trans}}^{j\rightarrow r}}(e)\) and \(\sum_{v_j\in p_j\backslash \{v_j\}}{w_{\mathrm{block}}^{j\rightarrow r}}(v_j)\) are defined similarly but along the path from leaf node \(v_j\) to the root node \(v_r\).

Given any non-leaf node \(v_k\), two paths to the root node \(v_r\) that intersect at \(v_k\) can always be found. Hence, the synchronization delay along these two paths can be divided into two parts: the delay from the leaf nodes to the intersecting node \(v_k\), and the delay from the intersecting node \(v_k\) to the root node \(v_r\).
\begin{align}
\nonumber
w(p_i)&=\underset{\text{from the leaf node }v_i\text{ to the intersecting node }v_k}{\underbrace{\left( \sum_{e\in p_{i\rightarrow k}}{w_{\mathrm{trans}}^{i\rightarrow k}}(e)+\sum_{v\in p_{i\rightarrow k}\backslash \{v_i\}}{w_{\mathrm{block}}^{i\rightarrow k}}(v) \right) }}\\
&+\underset{\text{from the intersecting node }v_k\text{ to the root node }v_r}{\underbrace{\left( \sum_{e\in p_{k\rightarrow r}}{w_{\mathrm{trans}}^{k\rightarrow r}}(e)+\sum_{v\in p_{k\rightarrow r}\backslash \{v_k\}}{w_{\mathrm{block}}^{k\rightarrow r}}(v) \right) }},
\\
\nonumber
w(p_j)&=\underset{\text{from the leaf node }v_j\text{ to the intersecting node }v_k}{\underbrace{\left( \sum_{e\in p_{j\rightarrow k}}{w_{\mathrm{trans}}^{j\rightarrow k}}(e)+\sum_{v\in p_{j\rightarrow k}\backslash \{v_j\}}{w_{\mathrm{block}}^{j\rightarrow k}}(v) \right) }}\\
&+\underset{\text{from the intersecting node }v_k\text{ to the root node }v_r}{\underbrace{\left( \sum_{e\in p_{k\rightarrow r}}{w_{\mathrm{trans}}^{k\rightarrow r}}(e)+\sum_{v\in p_{k\rightarrow r}\backslash\{v_k\}}{w_{\mathrm{block}}^{k\rightarrow r}}(v) \right) }}.
\end{align}
Here, \(p_{i \rightarrow k}\) and \(p_{j \rightarrow k}\) denote the distinct paths from the leaf nodes \(v_i\) and \(v_j\) to the intersecting node \(v_k\), respectively. \(p_{k \rightarrow r}\) denotes the shared path from the intersecting node \(v_k\) to the root node \(v_r\). Given that \(p_{k \rightarrow r}\) is common, the cumulative parameter transfer and blockage delay along this path segment are identical.

For the sake of analysis, we assign the synchronization delay of the shared path \(p_{k \rightarrow r}\) as a constant \(c^{k\rightarrow r}\):
\begin{equation}
c^{k\rightarrow r}=\sum_{e\in p_{k\rightarrow r}}{w_{\mathrm{trans}}^{k\rightarrow r}}(e)+\sum_{v\in p_{k\rightarrow r}\backslash\{v_k\}}{w_{\mathrm{block}}^{k\rightarrow r}}(v).
\end{equation}

For the path segments from leaf nodes \(v_i, v_j\) to the intersecting node \(v_k\), we isolate the blockage delay at the intersecting node \(v_k\), thus the overall synchronization delay can be expressed as:
\begin{align}
\label{eq:sync-delay-i}\nonumber
w(p_i)=&\left( \sum_{e\in p_{i\rightarrow k}}{w_{\mathrm{trans}}^{i\rightarrow k}}(e)+\sum_{v\in p_{i\rightarrow k}\backslash \{v_i,v_k\}}{w_{\mathrm{block}}^{i\rightarrow k}}(v) \right) \\
&+w_{\mathrm{block}}^{i\rightarrow k}(v_k)+c^{k\rightarrow r},
\\
\label{eq:sync-delay-j}\nonumber
w(p_j)=&\left( \sum_{e\in p_{j\rightarrow k}}{w_{\mathrm{trans}}^{j\rightarrow k}}(e)+\sum_{v\in p_{j\rightarrow k}\backslash \{v_j,v_k\}}{w_{\mathrm{block}}^{j\rightarrow k}}(v) \right) \\
&+w_{\mathrm{block}}^{j\rightarrow k}(v_k)+c^{k\rightarrow r}.
\end{align}

Given the synchronization barrier at the root node, the overall synchronization delays for any two paths from the leaf nodes to the root node are identical. Therefore, in the above equations, we have \(w(p_i) = w(p_j)\).

For any intersecting node \(v_k\), there must exist a ``last child node'' that enables \(v_k\) to exit the synchronization barrier. Without loss of generality, assuming this child node is located on the path \(p_{i\rightarrow k}\), which implies that its blockage delay \(w_{\mathrm{block}}^{i\rightarrow k}(v_k) = 0\). Consequently, according to Eqs. \eqref{eq:sync-delay-i} and \eqref{eq:sync-delay-j}, the path segment \(p_{i\rightarrow k}\) should have the highest synchronization delay, or in other words, it is the slowest path.

In conclusion, for any non-leaf node \(v_k\), a path can be found from a leaf node to \(v_k\) that incurs zero blockage delay. By applying the recursion principle, a path from a leaf node to the root node can be identified, where every non-leaf node encountered on this path exhibits zero blockage delay. Without loss of generality, assuming this path originates at leaf node \(v_i\), the cumulative blockage delay on path \(p_i\) would be
\begin{equation}
\sum_{v_i\in p_i\backslash \{v_i\}}{w_{\mathrm{block}}^{i\rightarrow r}}(v_i)=0.
\end{equation}

Based on the above analysis, the unique path with zero blockage delay exhibits the following characteristics:
\begin{enumerate}
    \item \textit{It has the same synchronization delay as any other path from a leaf node to the root node;}
    \item \textit{When calculating the synchronization delay on this path, the blockage delay term equals zero and can be neglected;}
    \item \textit{As a trade-off, this path has the highest parameter transfer delay, denoted by $\sum_{e\in p_i}{w_{\mathrm{trans}}^{i\rightarrow r}}(e)$.}
\end{enumerate}

Hence, we use this unique path to define the synchronization delay of the given topology. Specifically, among all possible paths from leaf nodes to the root node, the one with the highest cumulative parameter transfer delay is selected. This delay is then used to define the synchronization delay, as formulated in Eq. \eqref{eq:tree-completion-time}.
\end{proof}

\section{Proof of Proposition \ref{prop:measure-accuracy}}
\begin{proof}
To accurately measure link throughput, it is necessary to understand what constitutes transmission latency. In network transmission, the total latency consists of two parts: the transfer latency at the end nodes and the propagation latency along the physical link. These two types of latency combine to give a total latency \( t_{\mathrm{true}} = t_r - t_s \), where \( t_{\mathrm{true}} \) is the target latency, \( t_r \) is the timestamp when the receiver completes data reception, and \( t_s \) is the timestamp when the sender starts to transmit data.

Let us first consider a measurement approach based on round-trip delay, e.g., the one used in \cite{zhou2021tsengine}. In this work, the sender \( s \) records the timestamp \( t_s \) when a model chunk is sent and the timestamp \( \bar{t}_s \) when an ACK is received. Then, the link throughput \( \tau \) can be estimated by:
\begin{equation}
\tau =\frac{S}{\left( {\bar{t}_{s}}-t_{\mathrm{s}} \right) /2}=\frac{S}{t_{\mathrm{true}}+\frac{t_{\mathrm{prop}}}{2}}
\end{equation}
where \( \bar{t}_{s} - t_s \) includes the round-trip transfer and propagation latencies. However, this approach incurs an error of \( \frac{t_{\mathrm{prop}}}{2} \), because the transfer latency of the return ACK is near-zero, while its propagation latency is comparable to that of the outgoing model chunk. Here, \( t_{\mathrm{prop}} \) denotes the propagation latency of the return ACK.

Next, we consider our one-way delay approach, where the sender \( s \) records the timestamp \( t_s \) when a packet is sent, and the receiver \( r \) records the timestamp \( t_r \) when the reception is completed. In this way, the link throughput \( \tau \) is estimated by:
\begin{equation}
\tau =\frac{S}{t_{\mathrm{r}}-t_{\mathrm{s}}}=\frac{S}{t_{\mathrm{true}}}
\end{equation}
This approach eliminates the need for a return ACK, thus removing the error term \( \frac{t_{\mathrm{prop}}}{2} \) found in the round-trip delay approach and providing a more precise measurement of link throughput. Furthermore, it offers faster measurements.
\end{proof}

\end{document}